\documentclass[trackchanges]{aastex701}
\submitjournal{ApJ}

\begin{document}

\title{Optically Invisible Galaxies at Cosmic Noon and beyond with JWST/UNCOVER}

\author[0009-0002-2173-0953]{Pralay Biswas}
\affiliation{National Centre for Radio Astrophysics, Tata Institute of Fundamental Research, Post Bag 3, Ganeshkhind, Pune 411007, India}
\email{pbiswas@ncra.tifr.res.in}
\author[0009-0002-9692-3899]{Rashi Jain}
\affiliation{National Centre for Radio Astrophysics, Tata Institute of Fundamental Research, Post Bag 3, Ganeshkhind, Pune 411007, India}
\email{rjain@ncra.tifr.res.in}
\author[0000-0002-1345-7371]{Yogesh Wadadekar}
\affiliation{National Centre for Radio Astrophysics, Tata Institute of Fundamental Research, Post Bag 3, Ganeshkhind, Pune 411007, India}
\email{yogesh@ncra.tifr.res.in}

\begin{abstract}
The traditional selection bias in high-redshift galaxy surveys toward rest-frame ultraviolet emission has constrained our understanding of the high-redshift universe by systematically excluding optically faint (observer-frame) galaxies. Utilizing JWST UNCOVER and MegaScience data of the Abell 2744 cluster field, we identify 113 high-redshift ($z > 2$) and 94 low-redshift ($z < 2$) HST-dark galaxies using a red $1.50-3.56 \ \mu$m color criterion. Their physical properties were derived using multiwavelength photometry from 20 JWST/NIRCam and 7 HST filters. Unlike classical submillimeter and \textit{Spitzer}-selected HST-dark galaxies that primarily identify the most massive, dusty, highly star-forming galaxies, our study uncovers a moderately dusty population across a significantly broader range of stellar mass and star-formation rate. Leveraging gravitational magnification and ultra-deep JWST imaging, we found HST-dark galaxies with a stellar mass as low as $10^{7.5}$ at $z>2$. These galaxies have small sizes and follow the star-forming main sequence. Our analysis reveals a prominent peak in the Star Formation Rate Density at $z \approx 4.5$ ($10.14^{+7.10}_{-4.44} \times 10^{-3} \, M_{\odot} \, \text{yr}^{-1} \, \text{Mpc}^{-3}$). We also characterize a sub-population at $z < 2$ of Highly Extincted Low-Mass analogs with extremely low stellar masses (median $\log M_{\star}/M_{\odot} \approx 7.10$) and high dust extinction (median $A_V \approx 1.97$ mag). Our sample demonstrates the unique power of JWST to reveal this previously missing galaxy population and to provide a more complete census of galaxies in the high-z universe.
\end{abstract}


\section{Introduction} \label{sec: introduction}

Tracing the history of star formation and mass assembly across cosmic time remains one of the paramount goals of modern extragalactic astronomy. At redshifts of $z>3$, the rest-frame optical emission from galaxies redshifts entirely beyond the wavelength coverage of the Hubble Space Telescope (HST). Consequently, understanding of early star-forming galaxies, such as Lyman Break Galaxies (LBGs), has historically relied almost entirely on rest-frame ultraviolet (UV) observations. This severe UV bias has long been suspected to result in an incomplete census of galaxies in the early Universe. The presence of interstellar dust severely attenuates UV emission, rendering heavily obscured systems completely invisible to the high-redshift optical surveys in the observed frame. Similarly, massive quiescent galaxies, which have very little to no star formation and lack bright, young UV-emitting stars, also remain hidden from traditional optical dropout selection criteria for high-redshift galaxies, creating a critical gap in our understanding of the early universe.

Although extremely dust-obscured star-forming galaxies, such as submillimeter galaxies (SMGs; having $870 \ \mu m$ flux $S_{870 \ \rm \mu m} > 6.2 \ \rm mJy$ \citet{Simpson2020}), have long been known to exist at $z>4$, they remain rare. In fact, classical SMGs are roughly 100 times less common than LBGs, with a very low sky density of only $\sim0.01$ arcmin$^{-2}$ \citep{Riechers2013, Marrone2018}. Because of this rarity, these classical SMGs contribute only a limited fraction \citep[$10-15\%$;][]{Barger2014, Michalowski2017} to the integrated cosmic star formation rate density (SFRD). Similarly, while massive quiescent galaxies have been identified up to $z\sim3-5$, they have also historically represented a notably rare population at these early epochs \citep{Tanaka2019, Carnall2020, Santucci2020, Valentino2020, Leung2026, Stevenson2026}. Consequently, it is highly likely that early cosmic star formation and stellar mass buildup are not actually dominated by these rare extremes, but rather by more typical dusty star-forming galaxies (DSFGs) that evaded traditional rest-frame UV selection.

The historical effort to uncover these optically invisible ``HST-dark'' galaxies relied heavily on combining HST optical data with near-infrared imaging from the Infrared Array Camera (IRAC) on the \textit{Spitzer} Space Telescope \citep{Huang2011, Caputi2015, Stefanon2015, Wang2016, Wang2019b, Franco2018, Alcalde2019, Fudamoto2021}. However, these early studies were severely limited by \textit{Spitzer}/IRAC's low sensitivity and broad point-spread function (PSF), which frequently led to severe blending and source confusion in crowded deep fields, complicating accurate photometry. Also, the limited availability of photometric data, with only a few broad bands, makes spectral energy distribution (SED) modeling via stellar population synthesis (SPS) and redshift estimation uncertain. The advent of the James Webb Space Telescope (JWST) has fundamentally transformed this observational landscape. With its unparalleled sensitivity, multi-band broad and medium filter coverage in the near-infrared (spanning $\lambda=1-5 \ \mu$m), and high spatial resolution, the JWST Near-Infrared Camera \citep[JWST/NIRCam;][]{Rieke2003, Rieke2005, Greene2017, Rieke2023} allows us to resolve and identify these red, optically dark sources to much fainter limits.

Early JWST photometric analyses of these sources, using data from the Cosmic Evolution Early Release Science Survey \citep[CEERS;][]{Finkelstein2025}, revealed a fundamental ``triality'' in the nature of HST-dark galaxies \citep{perez-gonzalez2023}. Based on their spectral energy distributions (SEDs), these extremely red sources could be one of three distinct populations: massive, dusty star-forming galaxies at $z\sim2-6$; massive quiescent galaxies at $z\sim3-5$; or strong young compact starbursts with high equivalent width emission lines at $6<z<7$. Degeneracies arising from the restriction to photometric data made it extremely difficult to distinguish between these scenarios in the absence of spectroscopic data. However, recent dedicated spectroscopic follow-up with JWST Near Infrared Spectrograph \citep[JWST/NIRSpec;][]{Jakobsen2022, Boker2023} has successfully broken this degeneracy. As demonstrated by \citet{Barrufet2025}, spectroscopy of these extremely red galaxies at $z>3$ confirms that the population is a diverse mix; while many are heavily dust-obscured and star-forming galaxies with wide range of $\rm H\alpha$ equivalent widths ($\rm 68 \AA < EW_{H\alpha} < 550 \AA$), a substantial fraction ($13^{+9}_{-6}\%$) are indeed quiescent galaxies, proving that these massive systems were systematically missed by earlier optical samples.

Crucially, the dusty star-forming galaxies within the ``HST-dark'' population differ fundamentally from classical submillimeter galaxies. Analysis by \citet{barrufet2023} has shown that, unlike classical SMGs, which typically reside well above the star-forming Main Sequence (SFMS) due to their high mass and bursts of star formation, these HST-dark galaxies are predominantly less massive systems with relatively lower star formation rates (SFRs), placing them on the SFMS. Their inclusion in the cosmic census is necessary to achieve an accurate understanding of galaxy evolution; they represent a highly significant contribution to the galaxy mass function and could substantially contribute to the obscured SFRD at early epochs. Furthermore, recent findings of even redder systems, such as ``NIRCam-dark" galaxies that appear only in JWST Mid-Infrared Instrument \citep[JWST/MIRI;][]{Glasse2015, Wright2023} observations \citep[e.g., the source ``Cerberus'';][]{Perez-Gonzalez2024}, indicate that the early Universe contains heavily obscured populations that are missing even in the deepest near-infrared surveys with JWST/NIRCam.

While recent JWST wide-field surveys have successfully characterized the bright and high-mass end of the HST-dark population \citep{barrufet2023, perez-gonzalez2023, Barrufet2025, Gentile2025}, understanding the complete evolutionary picture requires extending our census to the intrinsically fainter, lower-mass, more numerous galaxies that better represent the typical building blocks of the Universe. The present study addresses this critical gap by utilizing ultra-deep NIRCam imaging from the JWST Ultradeep NIRSpec and NIRCam ObserVations before the Epoch of Reionization \cite[UNCOVER;][]{Bezanson2024} survey, deliberately capitalizing on the gravitational lensing magnification provided by the massive foreground cluster Abell 2744. By probing the magnified galaxies behind this lensing cluster, we aim to uncover the intrinsically faint, lower-mass counterparts to the recently discovered massive HST-dark galaxies in unlensed JWST fields. Using gravitational lensing, we aim to determine whether the highly dust-obscured, main-sequence nature of these sources extends to the fainter galaxies, thereby providing an essential, hitherto missing piece to the puzzle of cosmic stellar mass assembly.

This paper is organized as follows. Section \ref{sec: Data} describes the observational data from the UNCOVER survey and our sample selection. Section \ref{sec: SPS} outlines the SED modeling with \texttt{BAGPIPES}. We describe the low redshift ($z<2$) ``HST-dark'' galaxies in Section \ref{sec: Lowz}. In Section \ref{sec: Results}, we present our results, including the variety of physical properties and morphologies, compare our galaxy sample to SMGs and bright HST-dark sources, and discuss their contribution to the cosmic SFRD. We summarise and conclude the paper in Section \ref{sec: Summary}. Magnitudes are given in the AB system \citep{Oke1983}, and where necessary, we adopt a standard $\Lambda$CDM cosmology with $H_0 = 67.4 \rm \ kms^{-1}Mpc^{-1}$, $\Omega_m = 0.315$, and $\Omega_{\Lambda}$ = 0.685 \citep{Planck2020}.

\section{Data and Sample Selection} \label{sec: Data}

\begin{figure}
    \centering
    \includegraphics[width = 0.7\textwidth]{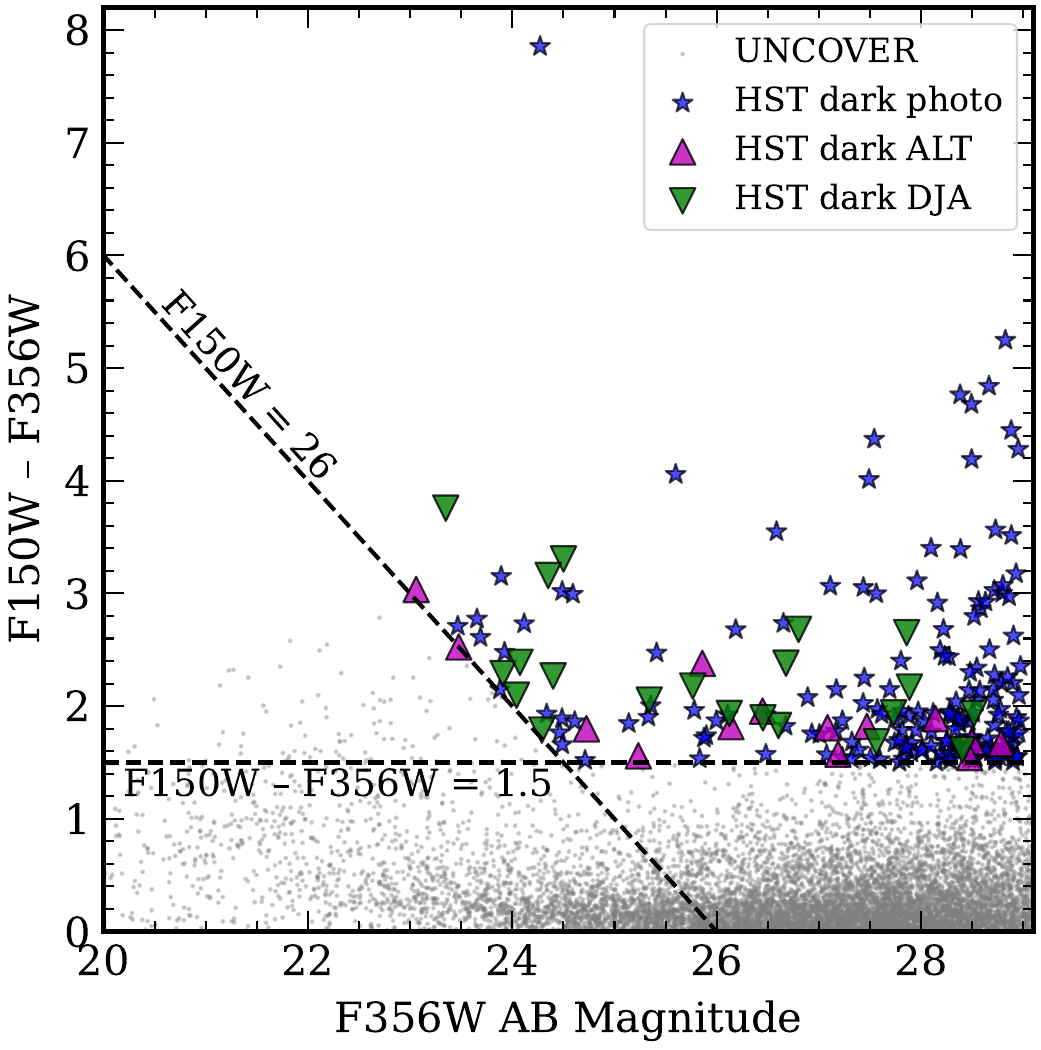}
    \caption{color-magnitude (F150W--F356W vs F356W) diagram to select HST-dark galaxies. The black dashed lines show the color cut F150W $-$ F356W = 1.5 and the F150W = 26 mag limit. The grey points represent the entire UNCOVER sample from the ``SUPER" catalog. The blue stars represent the selected HST-dark galaxies without spectroscopic data. The magenta upper triangle and green lower triangle, respectively, represent HST-dark galaxies with spectroscopic data from the ALT and DJA.}
    \label{fig: selection}
\end{figure}

Our analysis utilizes ultra-deep imaging of the Abell 2744 lensing cluster ($z=0.308$) from the UNCOVER \citep{Bezanson2024} treasury survey and the MegaScience medium-band survey \citep{Suess2024} taken from the UNCOVER data release 3 (DR3). UNCOVER survey (JWST Cycle 1) provides deep imaging over $\sim45$ arcmin$^2$ in seven NIRCam filters, including six broad-band filters (F115W, F150W, F200W, F277W, F356W, and F444W) and one medium-band filter (F410M). These observations reach $5\sigma$ limiting magnitudes of $\sim 29-30$ AB mag, providing the sensitive long-wavelength ``anchor" necessary to detect the dusty continuum of HST-dark galaxies. The MegaScience survey (JWST Cycle 2) builds upon the UNCOVER survey by contributing the remaining NIRCam filter set, including the two bluest broad-band filters (F070W and F090W), and 11 additional medium-band filters (F140M, F162M, F182M, F210M, F250M, F300M, F335M, F360M, F430M, F460M, and F480M). The $5\sigma$ depth of the MegaScience survey ranges from $\sim 27-30$ AB mag. Together, these programs provide a continuous sampling of the spectral energy distribution (SED) from $0.7$ $\mu$m to $5.0$ $\mu$m. The inclusion of the medium bands is critical to our analysis, as they enable precise identification of strong emission lines (e.g., $H\alpha$, $H\beta$, and $[OIII]$) and more accurate characterization of the spectral energy distribution. This continuous sampling of the SED helps to significantly reduce the uncertainties in the stellar population modeling parameters and, most importantly, in the redshift \citep{Suess2024}. We have used data from the UNCOVER DR3\footnote{\url{https://jwst-uncover.github.io/DR3.html}}, which includes other JWST observations apart from UNCOVER (GO-2561, PI: Labbe \& Bezanson) and MegaScience (GO-4111, PI: Suess), like MAGNIF (GO-2883, PI: Sun), ALT (GO-3516; PI: Naidu \& Matthee), GO-3538 (PI: Iani), ERS-1324 (GLASS), and DD-2767. We also utilise the re-processed Hubble Space Telescope (HST) mosaics and photometry provided by the UNCOVER team \citep{Weaver2024}, which include photometry from 3 ACS/WFC (F435W, F606W, and F814W) and 4 WFC3/IR (F105W, F125W, F140W, and F160W) filters from \#11689 (PI: Dupke), \#13386 (PI: Rodney), \#13495 (PI: Lotz / HFF), \#13389 (PI: Siana), \#15117 (PI: Steinhardt / BUFFALO), \#17231 (PI: Treu), \#13495 (PI: Lotz / HFF), and \#15117 (PI: Steinhardt / BUFFALO).

The significant gravitational lensing magnification provided by the Abell 2744 cluster in the UNCOVER survey allows us to observe galaxies that are intrinsically $1.5–4$ times fainter than those found in other deep JWST surveys such as the CEERS \citep{Finkelstein2025}. The combination of deep broad and medium-band photometry and lensing magnification provides the most complete census of optically dark galaxies at cosmic noon and beyond.

We identify our sample using a JWST-only color-magnitude criterion applied to the UNCOVER+MegaScience photometric ``SUPER" catalogue\footnote{\url{https://jwst-uncover.github.io/DR3.html\#PhotometricCatalogs}} \citep[hereafter SUPER catalogue;][]{Weaver2024} taken from the UNCOVER DR3. The ``SUPER" catalogue provides photometry from optimally selected color apertures covering all JWST imaging available in the Abell 2744 field over an area of $\sim 56 \ \rm arcmin^2$. It also provides identification flags for the stars and artifacts. The catalog is based on F444W PSF-matched imaging in which the bright cluster galaxies and the intra-cluster light have been modeled and subtracted. In addition, we have used the \texttt{use\_phot} flag provided in the catalog. This is a SUPER flag which separates robust sources (\texttt{use\_phot} = 1) from the marginally detected sources, stars, artifacts, and objects with no usable photometry (\texttt{use\_phot} = 0). Our primary selection requires:$$\text{color criterion, } (\rm F150W - F356W) > 1.5 \text{mag}$$$$\text{HST faintness criterion, } \rm F150W > 26 \text{ mag}$$$$\text{NIRCam detection criterion, } \rm F356W \le 29 \text{ mag}$$. The $\rm F150W - F356W$ color we use serves as a proxy for the $\rm H - [4.5]$ color used in legacy H-dropout studies using \textit{Spitzer}/IRAC \citep{Caputi2015, Wang2016, perez-gonzalez2023, Barrufet2025}, while benefiting from the high sensitivity of the JWST. Importantly, our use of a faint limit as $F356W \le 29$ allows us to probe a lower-mass population of galaxies ($\log M_{\star}/M_{\odot} \le 9$) and to exclude very faint galaxies below the survey limit. Using these selection criteria, we found $662$ galaxies. We visually inspected multi-band cutouts of all these sources twice and removed residual artifacts, diffraction spikes, noise peaks, and objects detected in only one band to obtain a robust sample. After this, our final sample includes 208 HST-dark galaxies (see Figure \ref{fig: selection}).

Some of our sample galaxies are likely to have archival JWST spectroscopic data. To check this, we cross-matched our final sample of HST-dark galaxies with the Dawn JWST Archive\footnote{\url{https://dawn-cph.github.io/dja/}} (hereafter DJA)  and All the Little  Things \citep{altzenodo} in the Abell 2744 \citep[hereafter ALT;][]{Naidu2024} survey. For cross-matching, we used a matching radius of $0.1$ arcsec. The DJA is a public JWST data repository. We have matched our final sample of 208 galaxies to the latest JWST/NIRSpec spectroscopic catalog (version 4.0) available in DJA. 27 galaxies from our sample have spectroscopic data in the DJA. We decided to use DJA spectroscopic redshifts only for galaxies with high-quality spectra (\texttt{grade} $=3$). After removing 5 galaxies\footnote{One of these galaxies is the UNCOVER-z13 \citep[MSA ID 13077;][]{Wang2023}.} with \texttt{grade} $<3$ spectra, we were left with 22 galaxies with DJA spectra in the Abell 2744 field. ALT is a JWST Cycle 2 survey that carried out NIRCam grism spectroscopy of small dwarf galaxies. ALT has spectroscopic redshifts measured for 1630 galaxies. After cross-matching our sample with the ALT catalog, we found 23 galaxies. There were 9 galaxies present in both ALT and DJA. For these 9 common galaxies, the redshifts from DJA and ALT are similar (with a maximum difference of 0.01), so we used the DJA redshifts. Therefore, using DJA and ALT data, we have a total of 36 HST-dark galaxies with known spectroscopic redshifts. All of these 36 galaxies have $z>2$.

\section{Stellar Population Synthesis (SPS) and redshift distribution} \label{sec: SPS}
 
\begin{figure}
    \centering
    \includegraphics[width = 0.5\textwidth]{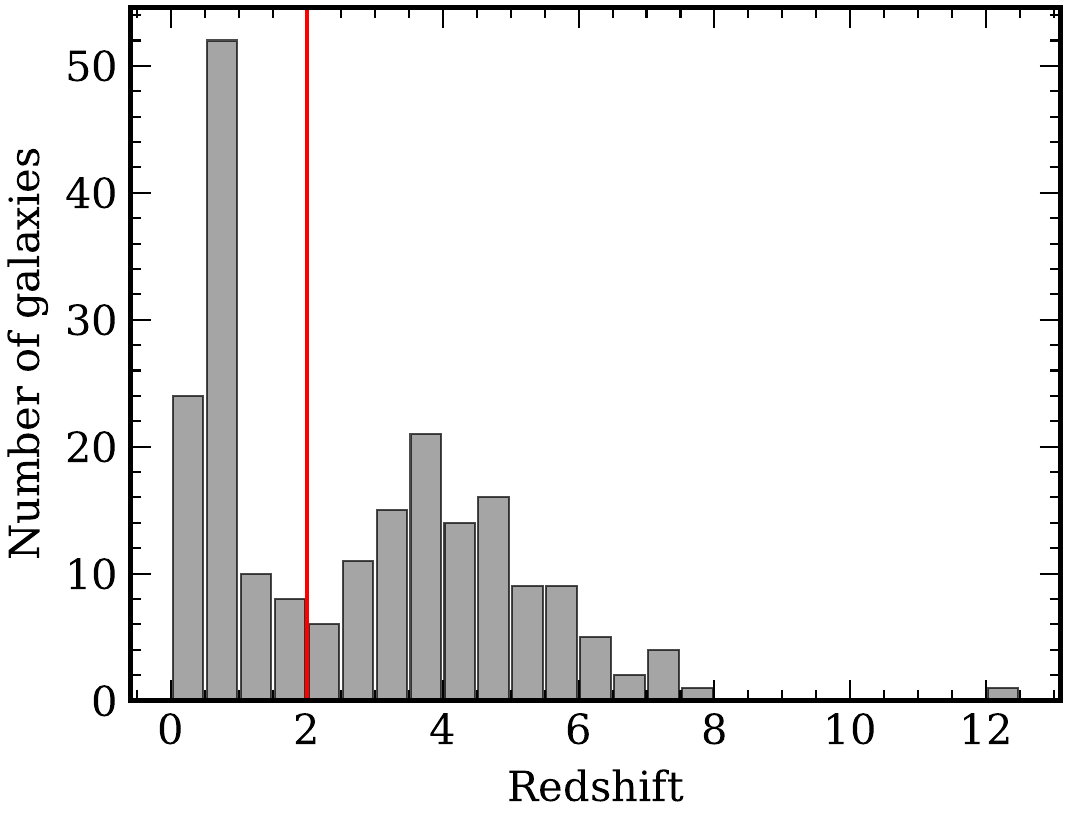}
    \caption{Distribution of redshift for the 208 HST-dark galaxies is shown. Along with the high redshift ($z>2$) HST-dark galaxies, we can also see a population of low redshift ($z<2$) HST-dark galaxies. We can also see one Lyman-break galaxy in our sample at $z=12.34$.}
    \label{fig: redshift_distribution}
\end{figure}

We use Bayesian Analysis of Galaxies for Physical Inference and Parameter EStimation \cite[\texttt{BAGPIPES};][]{Carnall2018, Carnall2019b} to estimate the physical properties of HST-dark galaxies. \texttt{BAGPIPES} is a Python program based on the Bayesian framework for SPS modeling of galaxies. It rapidly generates complex model galaxy spectra from the far-UV to the microwave by synthesizing stellar populations, dust attenuation, and nebular emission, and can be fitted to a combination of observed photometric and spectroscopic data.

For this analysis, we employ the 2016 version of \citet{Bruzual2003} SPS models, which uses the Medium-resolution Isaac Newton Telescope library of empirical spectra \cite[MILES;][]{Falcon-Barroso2011} for the UV-optical region of the spectrum. The models are implemented assuming a \citet{Kroupa2002} initial mass function. We model the star formation history (SFH) using the flexible delayed-$\tau$ model, allowing for a wide range of stellar ages and formation times. Given the likely dusty nature of  HST-dark galaxies, we model dust attenuation using the \citet{Calzetti2000} prescription, allowing the $V$-band attenuation ($\rm A_V$) to vary across a wide range of priors from 0 to 6. This high upper limit of 6 is necessary to capture the heavy dust obscuration characteristic of these galaxies. One of the critical components of our modeling is the inclusion of the nebular emission. \texttt{BAGPIPES} implements this by using the latest version of the Cloudy photoionization code \citep{Ferland2017} to self-consistently model the nebular continuum and line emission alongside the stellar continuum. We included nebular emission by varying the ionization parameter ($\log U$) from $-4$ to $-2$. The choice of \texttt{BAGPIPES} models, model parameters, and priors is summarised in Table \ref{tab:bagpipes_priors}. The posterior distributions of the parameters are sampled using the MultiNest nested sampling algorithm \citep{Feroz2008, Feroz2009, Feroz2019}, ensuring a robust exploration of the multi-dimensional parameter space and providing reliable uncertainties for the derived physical quantities.

We use 27-band multiwavelength photometry from the SUPER catalog in the UNCOVER DR3, as described in Section \ref{sec: Data}, for the SED modeling. We have also used the magnifications available from the UNCOVER DR4 SUPER magnifications catalog v2.0\footnote{\url{https://jwst-uncover.github.io/DR4.html\#UpdatedLensMag}}, which include magnifications for all the objects in the DR3 SUPER catalog \citep{Furtak2023a, Price2025} to demagnify the flux densities available in the SUPER catalog before providing them as inputs to \texttt{BAGPIPES}. An error of 5\% was added in quadrature with the uncertainty given in the SUPER catalog to account for systematic and calibration uncertainties, following the prescription of \citet{Boyer2022} and \citet{Wang2024}. For the 36 galaxies with spectroscopic redshift available from the DJA and ALT, we fixed the redshift for the SED fitting, and the other SPS parameters were the same as in Table \ref{tab:bagpipes_priors}.

Figure \ref{fig: redshift_distribution} shows the redshift distribution of the 208 HST-dark galaxies. Given the high redshift nature of the HST-dark galaxies, we expected our sample to be dominated by galaxies at $z>2$. However, we see a significant population of HST-dark galaxies (94 in number) at $z<2$. We also detect one Lyman break galaxy at $z=12.34$\footnote{This galaxy is known as the GHZ2/GLASS-z12 \citep{Naidu2022, Castellano2022, Castellano2024}.}. As Lyman break galaxies are fundamentally different from the HST-dark galaxies that are the focus of this paper, we have removed the $z=12.34$ galaxy from further study. We divide the full sample of HST-dark galaxies into two sub-samples: higher-redshift ($z>2$) and lower-redshift ($z<2$). This division at $z=2$ is not strictly robust, as many $z<2$ sources---and a few $z>2$ sources---have broad photometric redshift posteriors. This implies that some $z<2$ HST-dark galaxies may turn out to be high-redshift sources with more accurate redshift measurements (also see Section \ref{sec: Lowz}), while very few $z>2$ galaxies would similarly shift to $z<2$. The latter is less likely, since the redshift posteriors of $z>2$ sources are generally better constrained than those of the $z<2$ sample. However, there are 23 (out of 113) galaxies in the high-redshift HST-dark sample in which the lower 16th percentile reaches a redshift of 1.5 and has a broader redshift posterior. These galaxies are faint, with low stellar mass ($\rm \log M_* \lesssim 8.5 \ M_\odot$). Removing these galaxies slightly changes the median values of the sample. However, the main conclusions of our paper remain unchanged. Therefore, we have retained those galaxies in our sample. This paper primarily focuses on the HST-dark galaxy sample (113 galaxies) at $z>2$. HST-dark galaxies will refer to this primary sample unless explicitly mentioned otherwise. Before discussing the primary sample, we now briefly discuss the low-redshift ($z<2$) HST-dark galaxies in Section \ref{sec: Lowz}.

\begin{table}[ht]
\centering
\caption{Summary of \texttt{BAGPIPES} Model Parameters and Priors for the HST-dark Galaxy Sample.}
\label{tab:bagpipes_priors}
\begin{tabular}{l l l l}
\hline
\hline
Component & Parameter & Prior Distribution & Range/Value \\
\hline
\textbf{Stellar Population} & IMF & Fixed (Kroupa \& Boily 2002) & -- \\
\hline
\textbf{Star Formation History} & Model & Delayed $\tau$ & $t e^{-t/\tau}$ \\
& Age ($Gyr$) & Uniform & $[0.1, \text{Age of Universe}]$ \\
& $\tau$ ($Gyr$) & Uniform & $[0.01, 10]$ \\
& $\log(M_*/M_\odot)$ & Uniform & $[4, 15]$ \\
& Metallicity ($Z/Z_\odot$) & Logarithmic & $[0.01, 2.5]$ \\
\hline
\textbf{Dust Attenuation} & Model & Calzetti (2000) & -- \\
& $A_V$ (mag) & Uniform & $[0, 6]$ \\
\hline
\textbf{Nebular Emission} & $\log(U)$ & Uniform & $[-4, -2]$ \\
\hline
\textbf{Redshift} & $z$ & Uniform & $[0, 15]$ \\
\hline
\hline
\end{tabular}
\end{table}

\section{Low redshift ($z<2$) HST-dark Galaxies} \label{sec: Lowz}

\begin{figure}
    \centering
    \includegraphics[width = 0.85\textwidth]{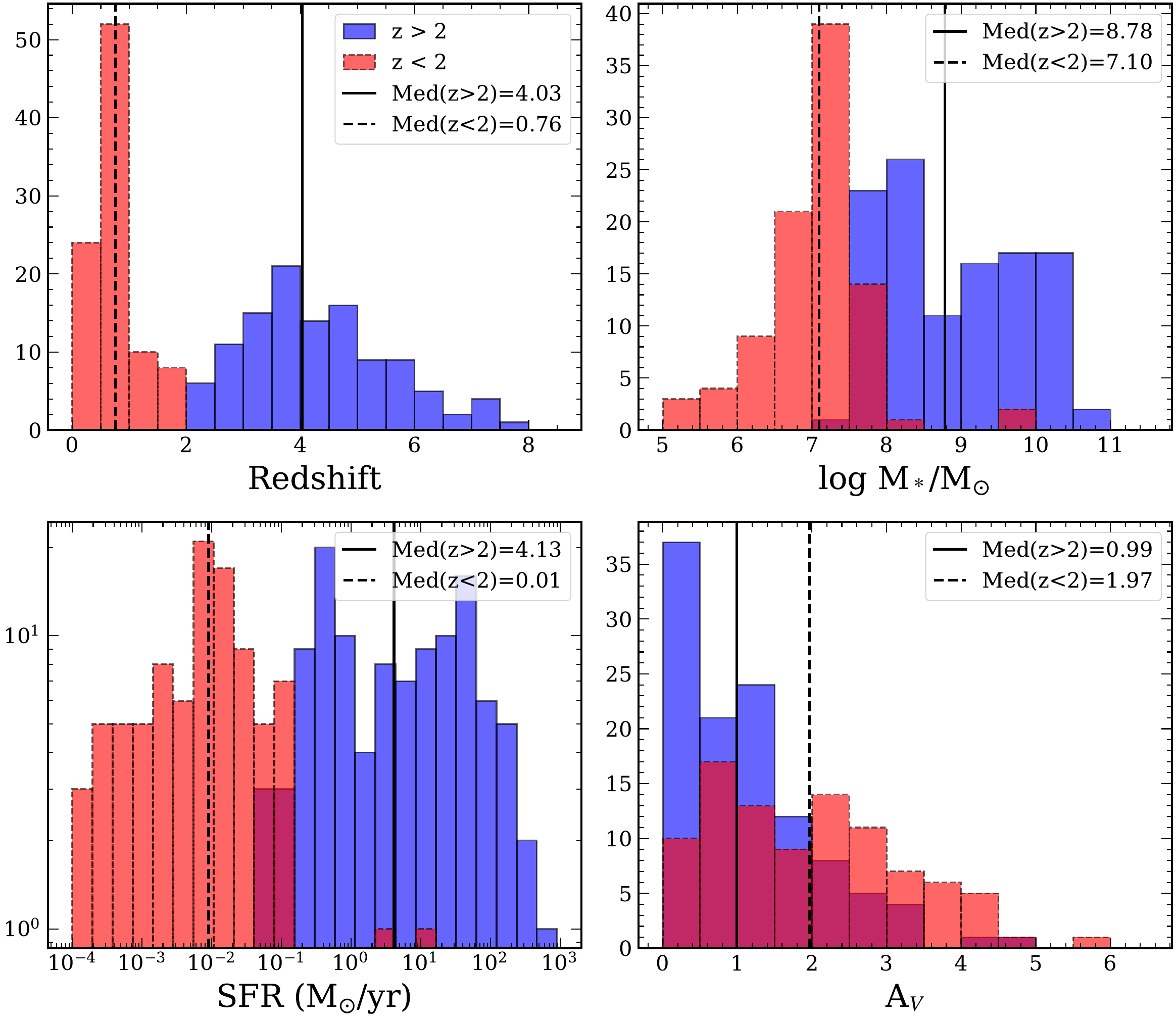}
    \caption{Histogram of the redshift (upper left), stellar mass (upper right), SFR (lower left), and dust attenuation (lower right) of HST-dark galaxies. HST-dark galaxies with $z<2$ are shown in red color and black dashed edge lines. The HST-dark galaxies with $z>2$ are shown in blue and with black edge lines. The respective median values for $z<2$ are shown with black dashed lines, and the same for $z>2$ are shown with black lines. We can see that $z<2$ HST-dark galaxies have lower masses, very low SFRs, and higher dust attenuation than $z>2$ HST-dark galaxies.}
    \label{fig: histogram_comb}
\end{figure}


\begin{figure}
    \centering
    \includegraphics[width = 0.7\textwidth]{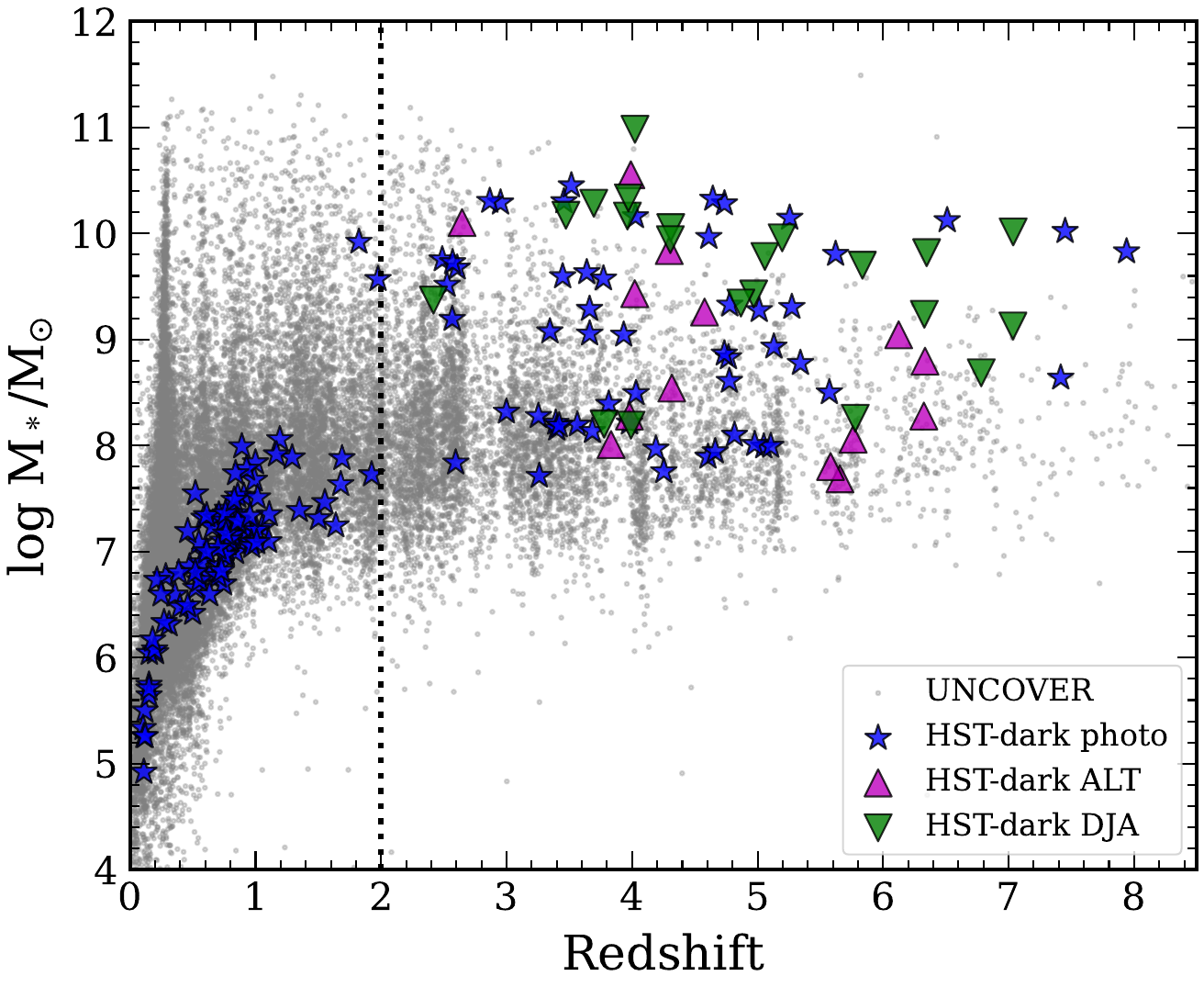}
    \caption{Distribution of stellar mass is shown with redshift. The grey points are the whole UNCOVER sample. The blue stars represent HST-dark galaxies with no spectroscopic data. The magenta upper triangle and the green lower triangle show HST-dark galaxies with spectroscopic data from ALT and DJA, respectively. The plot shows that the HST-dark galaxies with $z<2$ have lower stellar mass and are not a complete sample. The HST-dark galaxies with $z>2$ have higher stellar masses and are a complete sample.}
    \label{fig: redshift_mass}
\end{figure}

While the primary focus of our HST-dark selection is the high-redshift ($z > 2$) universe, our sample selection criteria also identify a significant sub-population of 94 galaxies residing at lower redshifts ($z < 2$). They exhibit a median redshift of $z \approx 0.76$ (ranging from $z \approx 0.1$ to $1.98$) and possess physical properties that deviate starkly from the standard scaling relations observed in the local universe.

See Figure \ref{fig: histogram_comb} (red color histogram) for the distribution of HST-dark galaxies with $z<2$ in redshift, stellar mass, SFR, and dust attenuation. This low-redshift sample is characterized by extremely low stellar masses and disproportionately high dust attenuation. The median $\rm A_V$ of the entire UNCOVER sample excluding the HST-dark galaxies within the same stellar mass range as the low-redshift ($z<2$) HST-dark galaxies is 0.26. The median stellar mass of the low-redshift HST-dark sample is $\log M_{\star}/M_{\odot} \approx 7.10$, placing these objects firmly in the dwarf galaxy regime. Despite their low masses, they exhibit a median dust extinction of $A_V \approx 1.97$ mag, with a maximum of $A_V \approx 5.71$ mag. These properties are remarkably similar to the Highly Extincted Low-Mass (HELM) galaxies identified by \citet{Bisigello2023, Bisigello2026}. Like the HELM population, our $z < 2$ sources show significantly higher dust attenuation and extremely low SFRs, with a median $\text{SFR} \approx 0.009 \, M_{\odot} \text{yr}^{-1}$.

\begin{figure}
    \centering
    \includegraphics[width = 0.7\textwidth]{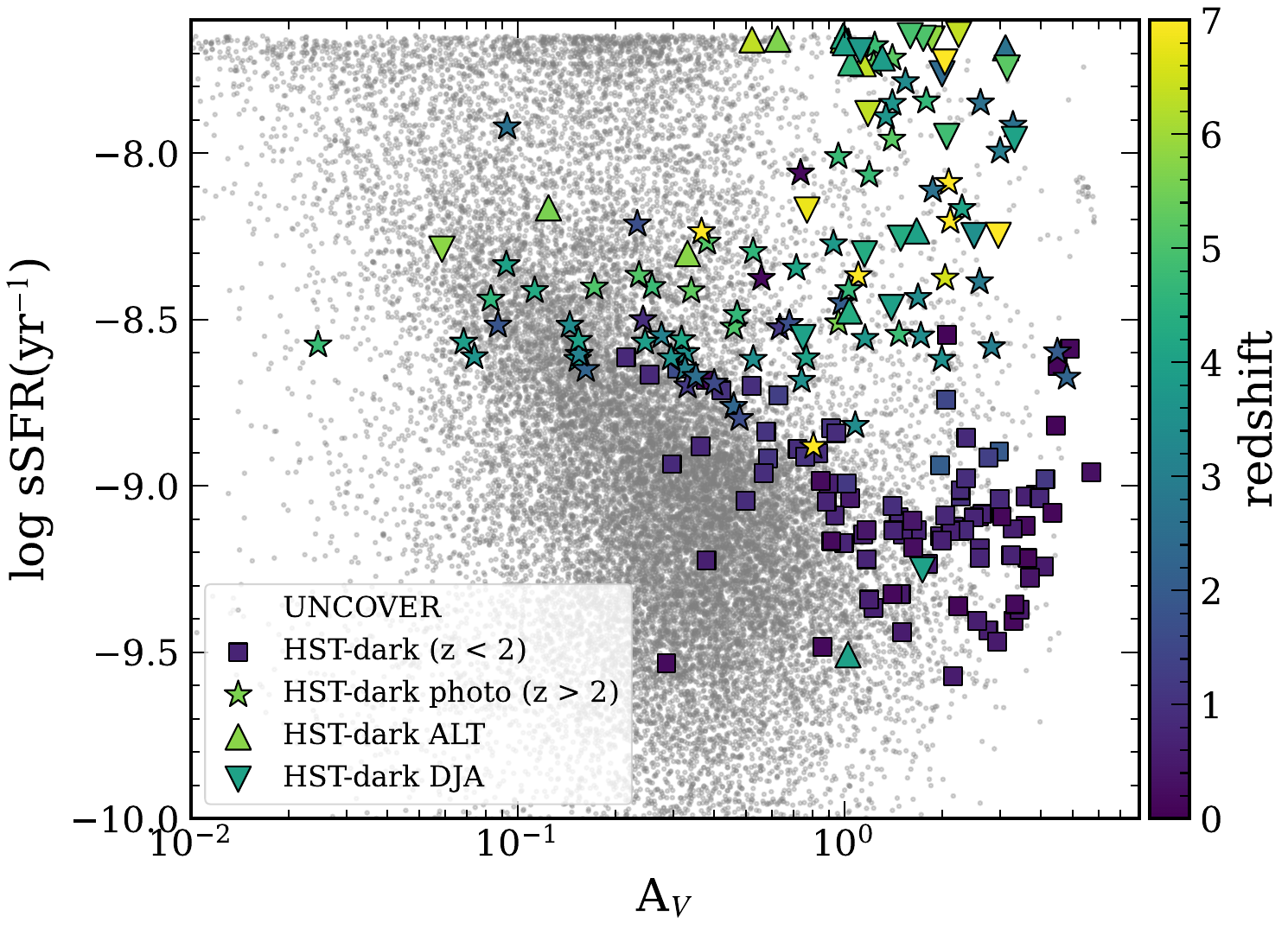}
    \caption{Distribution of HST-dark galaxies in the dust attenuation ($\rm A_V$) vs $\rm \log(sSFR(yr^{-1}))$ plane. Grey points represent all the UNCOVER survey galaxies. The stars, upper and lower triangle markers, represent the high-redshift ($z>2$) galaxies. The low-redshift ($z<2$) HST-dark galaxies are marked with squares. HST-dark galaxies are colored according to their redshift. HST-dark galaxies with $z<2$ have high dust content and lower sSFR than the HST-dark galaxies with $z>2$.}
    \label{fig: lowz_dust_ssfr}
\end{figure}

Figure \ref{fig: redshift_mass} shows redshift vs stellar mass for the whole sample. It is evident that these $z < 2$ galaxies have lower stellar mass and are only visible at lower redshifts, due to the Malmquist bias. We have also plotted our high dust attenuation sample in the $\log(sSFR)$ vs $A_V$ plane (Figure \ref{fig: lowz_dust_ssfr}). It shows that HST-dark galaxies with $z<2$ have systematically lower specific star formation rates (sSFR) and higher dust content. Therefore, the presence of these galaxies in our HST-dark catalog is a direct result of their intrinsic faintness, driven by a combination of low stellar mass and low SFR, further exacerbated by significant dust. These ``dusty dwarfs" are essentially the low-mass extension of the massive HST-dark galaxies, but this population can now be detected due to the extreme sensitivity of the JWST, albeit only at low redshifts.

Many of these $z < 2$ galaxies are properly detected in only 3 to 4 longer-wavelength bands of NIRCam. As a result, their SEDs are not properly constrained, and so they often exhibit broad photometric redshift posterior ($P(z)$) distributions. This redshift ambiguity also arises because the red, dusty continuum of a low-mass dwarf at $z \sim 0.8$ can closely mimic the spectral shape of the Lyman break in a high-redshift galaxy. This ``interloper" problem has been frequently encountered in recent JWST surveys, where dusty low-$z$ dwarfs are frequently misidentified as high-$z$ candidates before being classified correctly through more rigorous SED fitting or spectroscopic follow-up \citep{Zavala2023, Arrabal2023, Bisigello2023, Zavala2026}. The broad $P(z)$ distributions for our $z < 2$ sample suggest that while they are likely HELM-like analogs, a small fraction could potentially be higher-redshift contaminants, necessitating future NIRSpec observations to definitively break the age-dust-redshift degeneracy.

\section{Results and Discussion} \label{sec: Results}

\subsection{Properties of HST-dark Galaxies} \label{subsec: Properties}
 
With the unprecedented sensitivity of JWST/NIRCam, and by combining data in broad-band filters with the MegaScience medium-band filters, we can now accurately constrain the photometric redshift and the SEDs of the HST-dark galaxies. We can derive the physical parameters, such as stellar masses, SFR, and dust content, of these galaxies more reliably than the previous studies with IRAC-selected sources. We will now discuss the physical properties of the HST-dark galaxies derived from the \texttt{BAGPIPES} SED fitting (see Figure \ref{fig: histogram_comb} in blue).

Our analysis suggests that HST-dark galaxies are not a homogeneous sample and include diverse galaxies spanning a range of redshifts, stellar masses, and SFRs. By their very nature, the vast majority of HST-dark galaxies reside at redshifts $z>3$. Recent discoveries utilizing JWST have confirmed their presence well into the Epoch of Reionization at $z\sim6-8$ \citep{barrufet2023, perez-gonzalez2023}, whereas the previous surveys of HST-dark galaxies with \textit{Spitzer} were limited to $z<6$. The median redshift of our final sample of HST-dark galaxies is $4.03$. We have 12 galaxies with $z>6$ (see top left panel of Figure \ref{fig: histogram_comb} in blue). 


Our analysis confirms that the HST-dark population is far more diverse than the massive ($\rm M_{\star}>10^{10}M_{\odot}$), infrared-bright starbursts identified in early HST, \textit{Spitzer} surveys \citep{Wang2019b}. Moreover, the population encompasses a remarkably broad mass range. Previous JWST surveys reveal distributions spanning ($\rm \log M_{\star}/M_{\odot} \sim 9-11$) with median values around $\rm 10^{10} \ M_{\odot}$ \citep{barrufet2023, perez-gonzalez2023, Xiao2023, Barrufet2025}. Furthermore, JWST data have uncovered a population of High-Extinction, Low-Mass (HELM) galaxies similar to the HST-dark galaxies with $z<2$ in our sample that extend this distribution down to ($\rm \log M_{\star}/M_{\odot} \leq 8$), proving that strong dust obscuration is not exclusively tied to the highest-mass systems in the early Universe \citep{Bisigello2023, Bisigello2026}. While our sample includes massive systems ($\log M_{\star}/M_{\odot} \gtrsim 10$) consistent with those found by \citet{barrufet2023, perez-gonzalez2023}, it also reveals a significant population of lower-mass galaxies ($\log M_{\star}/M_{\odot} \lesssim 9.0$). Our sample resides in the regime of $\log M_{\star}/M_{\odot} \approx 7.5 - 10.5$ with a median value of $\rm M_{\star} = 10^{8.78} \ M_{\odot}$, representing a class of galaxies that were previously missing even to early JWST surveys like CEERS having somewhat poorer sensitivity than UNCOVER. \texttt{BAGPIPES} provides not only best-fit parameter values but also their full posteriors. Using the posteriors for individual galaxies, we constructed a mean posterior for the redshift and stellar mass of HST-dark galaxies and calculated the 50th-percentile values from these mean posteriors. These values are consistent with the median redshift and stellar mass of the primary sample (see Appendix \ref{sec: posterior}).

We find that these galaxies have a range of dust content, with 75\%  having V-band attenuation values $A_V > 0.3$ mag. In the SED modeling, we allowed $A_V$ to vary from 0 to 6. However, the dust attenuation reaches only up to 5, with only 2 galaxies having $A_V$ above 4, similar to \citet{barrufet2023}. For our sample, the median dust attenuation is 0.99 mag. Because of the extreme depth and lensing nature of the UNCOVER survey, we are, for the first time, seeing the low-mass, less-dusty population of HST-dark galaxies.

Our sample also exhibits a broad range of sSFRs, from weaker star-forming galaxies ($\text{sSFR} \sim -10.0 \ \text{yr}^{-1}$) to intense star-forming galaxies having $\text{sSFR} > -8.0 \, \text{yr}^{-1}$. Not all of the red sources are actively forming stars; high-resolution spectroscopy has revealed that a substantial fraction (roughly $13–18\%$) are actually massive quiescent galaxies that rapidly quenched very early \citep{perez-gonzalez2023, Barrufet2025}. More than $75\%$ of our HST-dark galaxies have $\text{SFR} < 10 \, M_{\odot} \text{yr}^{-1}$ with a median value of $4.13 \, M_{\odot} \text{yr}^{-1}$. The fact that these low-mass, dusty galaxies contribute so substantially to the overall stellar mass budget at $z > 3$ suggests that the ``missing'' part of the cosmic SFRD is not just in the most massive galaxies, but is spread across the mass function. This implies that the transition from ``UV-bright'' to ``dust-obscured'' occurs at lower stellar masses than previously predicted or seen, potentially requiring a revision of how we model feedback and metal enrichment in the first few billion years of cosmic time (see also \ref{subsec: SFRD}).


\subsection{Star Forming Main Sequence} \label{subsec: MS}

\begin{figure}
    \centering
    \includegraphics[width = 0.7\textwidth]{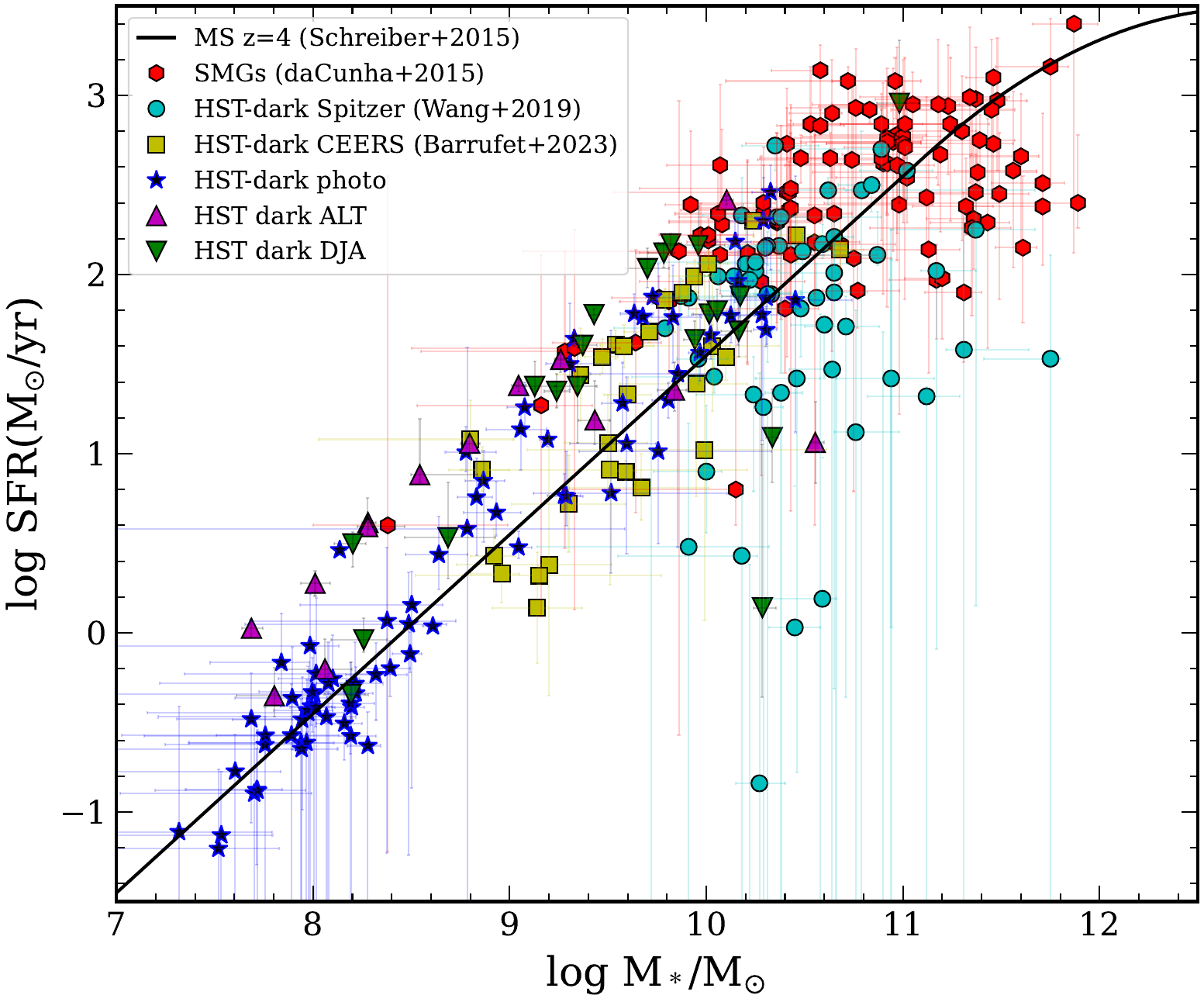}
    \caption{Star-forming main sequence (SFR against stellar mass) is shown. Our sample of HST-dark galaxies is shown with blue stars, magenta upper triangles, and green lower triangles. The HST-dark sample from \citet{barrufet2023} is shown with yellow squares. \textit{Spitzer}-selected optically dark galaxies from \citet{Wang2019b} are represented by cyan circles. SMGs are shown with red hexagons from \citet{daCunha2015}. Our sample of HST-dark galaxies probes a lower stellar mass and lower SFR region of the MS plot not seen before.}
    \label{fig: SFMS}
\end{figure}

The bulk of star-forming galaxies assemble their stellar mass through a secular, steady process, placing them on a tight correlation between SFR and stellar mass known as the Star-Forming Main Sequence (SFMS). Determining where optically faint or HST-dark galaxies lie relative to this sequence is crucial for understanding whether they represent normal galaxy growth or extreme, episodic events in the early Universe. To characterize the evolutionary state of our sample, we place our sample of HST-dark galaxies on the stellar mass versus SFR ($M_{\star}$–SFR) plane (see Figure \ref{fig: SFMS}). To compare our results with the previous studies of dusty galaxies, we have plotted the ALESS SMGs \citep{daCunha2015} and HST-dark galaxies selected from \textit{Spitzer} \citep{Wang2019b} and JWST \citep{barrufet2023}. Our sample spans a wide range of physical properties, with SFRs extending from relatively quiescent levels ($\sim 0.06 \, M_{\odot} \text{yr}^{-1}$) to extreme starbursts exceeding $\sim 900 \, M_{\odot} \text{yr}^{-1}$ (mean $\text{SFR} \approx 35.3 \, M_{\odot} \text{yr}^{-1}$). When compared to the established SFMS at $z \sim 4$ \citep{Schreiber2015}, the majority of our sources lie squarely within the expected scatter of the sequence, as also seen by \citet{barrufet2023}. This suggests that the ``optically dark'' nature of these galaxies is not necessarily a sign of exotic, off-sequence starburst activity, but rather a ubiquitous phase of typical galaxy growth that is simply obscured by dust.

The integration of UNCOVER ultra-deep imaging with MegaScience medium-bands enables us to trace the SFMS into a regime previously inaccessible to legacy surveys. The previous \textit{Spitzer}/IRAC and SMG studies \citep{daCunha2015, Wang2019b} were largely restricted to massive systems ($\log M_{\star}/M_{\odot} \gtrsim 10$). This demonstrates that previous surveys systematically missed the lower-mass end of the obscured main sequence and were heavily biased toward only the brightest, star-forming, and heavily dust-obscured galaxies. Even with earlier JWST surveys like CEERS, the HST-dark galaxy sample was limited to $\log M_{\star}/M_{\odot} \gtrsim 9$ \citep{barrufet2023, perez-gonzalez2023}. However, our sample probes nearly two orders of magnitude lower in stellar mass, from $10^{9} M_{\odot}$ down to a ``dusty dwarf'' regime at $\log M_{\star}/M_{\odot} \approx 7$. This significant extension to lower masses indicates that the HST-dark phenomenon is prevalent even in small-mass galaxies at early cosmic epochs.

Comparing our findings to classic Submillimeter Galaxies (SMGs), which typically represent the high-SFR (SFR $\sim 10^{3} M_{\odot} \rm yr^{-1}$) outliers of the SFMS, we find that our UNCOVER-selected sample captures the dust-obscured population with upto four orders of magnitude lower stellar mass and SFR. By accounting for these sources across the $z \approx 2-8$ range, we provide a more complete census of the stellar mass (up to $\log(M_*/M_\odot) \sim 7.0$) and SFR budget over this redshift range.

The identification of this previously missing low-mass population is driven by two key factors. First, the deeper effective sensitivity of the UNCOVER survey allows us to see intrinsically fainter galaxies. By leveraging deeper observations and gravitational magnification, our lowest-mass sources have intrinsic NIRCam magnitudes well below the typical detection limits ($F356W \gtrsim 28.5$ mag) of blank-field JWST surveys like CEERS \citep{Finkelstein2025} or COSMOS-Web \citep{Casey2023}. As a result, these ``dusty dwarfs'' would simply remain undetected in shallower surveys of JWST. Second, the broad photometric coverage of the MegaScience medium-band filters plays a crucial role. In broadband-only surveys, the SEDs of faint dusty galaxies are often poorly constrained. The medium bands provide a finely sampled SED that robustly determines spectral features. This improves the joint constraints on redshift, stellar mass, and dust attenuation, preventing these faint sources from being misclassified during SED modeling. Furthermore, the reliability of our sample is strongly supported by spectroscopy; out of our 113 high-redshift ($z>2$) HST-dark galaxies, we have spectroscopic redshifts available for 36 of them from the DJA and ALT.

\subsection{Dust attenuation} \label{subsec: Dust}
 
A defining characteristic of the HST-dark population is the significant attenuation of rest-frame ultraviolet and optical light by interstellar dust. In our sample, we find a broad distribution of V-band attenuation, ranging from nearly transparent ($A_V \approx 0.02$) to heavily obscured ($A_V \approx 4.80$), with a mean value of $A_V = 1.14$ mag. Interestingly, we find that the galaxies at the lower end of the dust distribution are characteristically low in both stellar mass and SFR. These sources are HST-dark, not due to extreme dust obscuration, but because even moderate dust columns compound their intrinsic faintness from low mass and low SFR. This combination of factors pushes them below the HST detection limit.

\begin{figure}
    \centering
    \includegraphics[width = 0.7\textwidth]{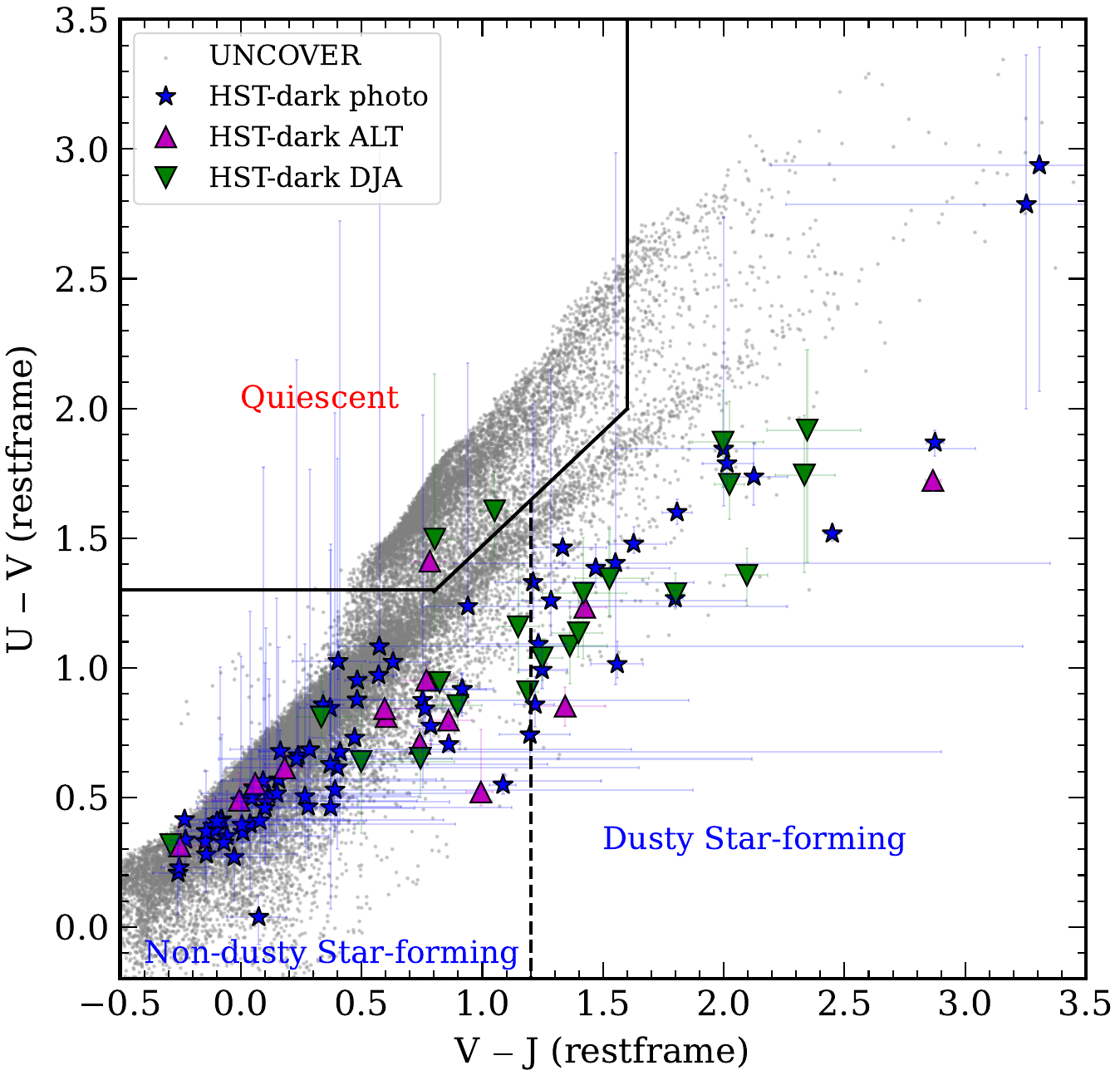}
    \caption{The UV vs VJ rest-frame color-color diagram is shown. Our sample of HST-dark galaxies is shown with blue stars, magenta upper triangles, and green lower triangles. All the UNCOVER galaxies are shown with grey points. The black dashed and solid lines are from \citet{Spitler2014} separating quiescent, dusty, and non-dusty star-forming galaxies. Our sample spans a range of UV and VJ colors, with almost all galaxies being star-forming. We also have three quiescent galaxies in our sample. However, given the uncertainties of the colors, there could be more galaxies in the quiescent region. The spectrum of one of the quiescent galaxies is shown in Appendix~\ref{sec: spectrum}.}
    \label{fig: uvj}
\end{figure}

To distinguish between dusty star formation and quiescence across this diverse sample, we utilize the rest-frame $U-V$ versus $V-J$ (UVJ) color-color diagnostic (see Figure \ref{fig: uvj}). The UVJ diagram is critical to break the degeneracy between age and dust. Our sample primarily occupies the dusty and non-dusty star-forming regions, similar to the sample of \citet{barrufet2023}. The galaxies with both low ($<0.5$) $U-V$ and $V-J$ were not previously seen \citep{barrufet2023, perez-gonzalez2023}. We can also see three galaxies in the quiescent region of the UVJ diagram, and we have spectroscopic redshifts for all of them (3.69, 3.96, and 3.99), one from ALT and two from DJA. Their existence is consistent with the claim of \citet{perez-gonzalez2023} that quiescent galaxies at $z<3$ do not satisfy the HST-dark color cut. The spectrum of one of the quiescent galaxies is shown in the Appendix \ref{sec: spectrum}. We can see that the spectrum is red, with no lines except for a weak $H{\alpha}$ line \citep[also see][]{Setton2024}. This confirms that the ``HST-dark" nature of our targets is predominantly driven by a combination of SFR and dust reddening. This broad distribution in $A_V$ and UVJ color within our sample suggests a diverse range of dust geometries and birth-cloud environments across the $z \approx 2$ to $8$ range.

\begin{table}[ht]
\centering
\caption{Comparison of median redshift and dust attenuation.}
\label{tab:sample_comparison}
\begin{tabular}{l c c c}
\hline
\hline
Sample Source & Median Redshift ($z$) & Median Attenuation ($A_{V}$) \\
\hline
HST-dark (UNCOVER; this work) & 4.03 & 0.99 \\
HST-dark \citep[CEERS;][]{barrufet2023} & 4.41 & 2.06 \\
SMGs \citep{daCunha2015} & 2.83 & 2.10 \\
\hline
\hline
\end{tabular}
\end{table}

\begin{figure}
    \centering
    \includegraphics[width = 0.7\textwidth]{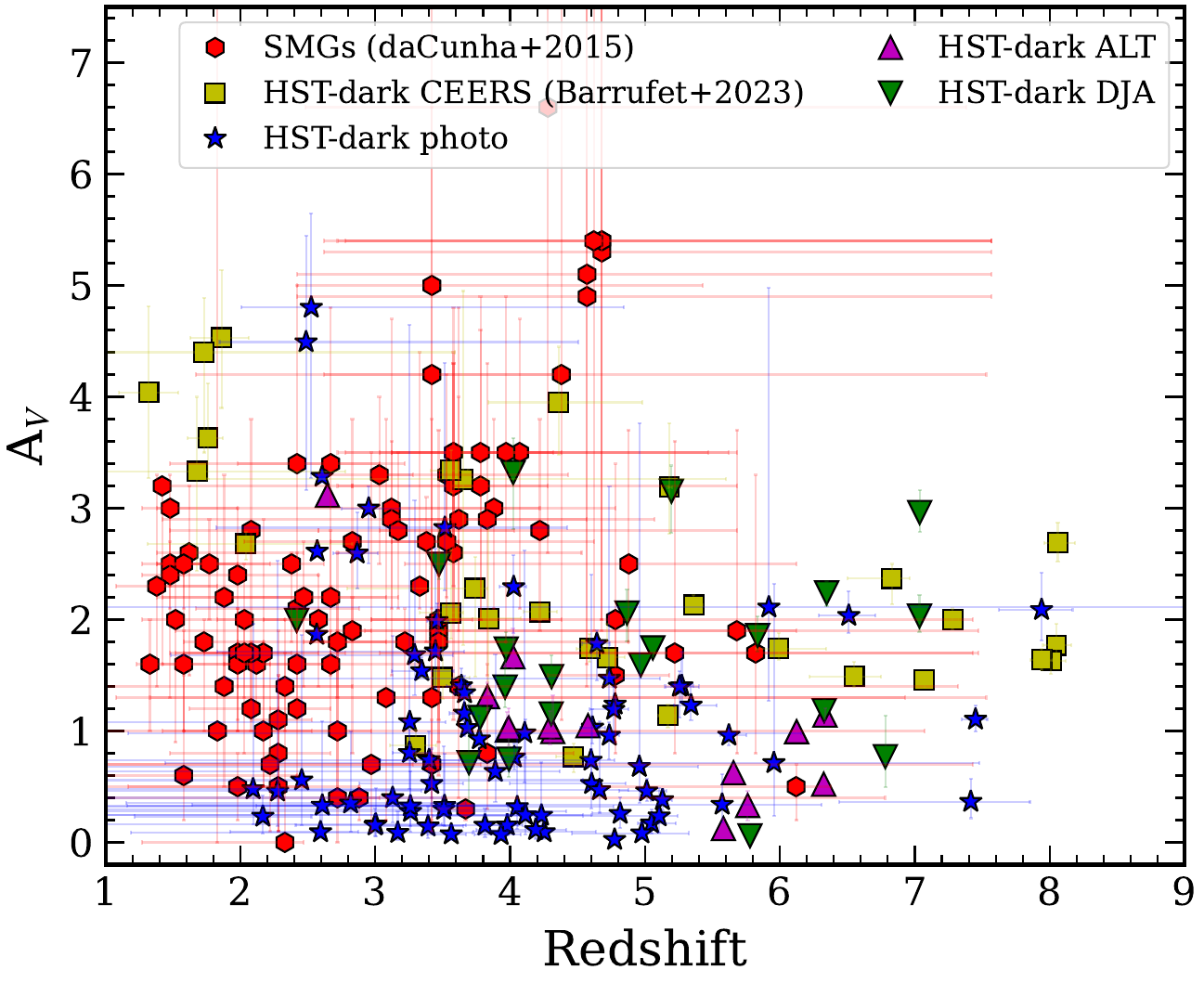}
    \caption{Dust attenuation ($A_V$) against redshift is represented. As in the previous figures, our sample of HST-dark galaxies is shown with blue stars, magenta upper triangles, and green lower triangles. SMGs are shown with red hexagons \citep{daCunha2015}. HST-dark galaxies from CEERS are shown with yellow squares \citep{barrufet2023}. HST-dark galaxies, on average, have lower dust content and are at a higher redshift than SMGs. Our sample probes a lower $A_V$ compared to both SMGs and HST-dark galaxies from CEERS.}
    \label{fig: dust_redshift}
\end{figure}

When comparing our results with SMGs and previous JWST studies, we see a clear hierarchy in the optically dark galaxies (see Figure \ref{fig: dust_redshift} and Table \ref{tab:sample_comparison}). Our sample sits at a significantly higher redshift (median $z \approx 4.0$) and lower dust attenuation (median $A_V \approx 1.0$) compared to classical SMGs (median $z \approx 3.0$ and $A_V \approx 2.0$). Our sample reaches similar high-redshift territory as the \citet{barrufet2023} sample (median $z \approx 4.4$ and $A_V \approx 2.0$) but with a notably lower $A_V$.

This highlights the unique discovery space of the UNCOVER survey. While \citet{barrufet2023} primarily identified the ``classical'' HST-dark galaxies (massive and dusty), our sample allows us to detect a ``fainter'' dark population. These galaxies are HST-dark not necessarily because they possess extreme dust ($A_V > 1$), but because their moderate dust content is sufficient to push their already faint, high-redshift flux below the detection limits of HST. This moderately dusty population, missing from both SMG and previous HST-dark samples, represents a crucial link in understanding the transition from the UV-bright to the dust-obscured phase of galaxy evolution.

\subsection{Size-Mass Relation} \label{subsec: SM}

\begin{table}[ht]
\centering
\caption{Median $r_{80}$ for the HST-dark and UNCOVER Populations ($3 < z < 5$).}
\begin{tabular}{c c c c c c c}
\hline
\hline
Mass Bin [log($\frac{M_{\star}}{M_{\odot}}$)] & $N_{\text{UNCOVER}}$ & Median $r_{80, \text{UNCOVER}}$ [kpc] & $N_{\text{HST-dark}}$ & Median $r_{80, \text{HST-dark}}$ [kpc] \\
\hline
$7-8$   & 833 & $1.47 \pm 0.03$ & 9  & $1.30 \pm 0.17$ \\
$8-9$   & 870 & $2.10 \pm 0.03$ & 16 & $1.57 \pm 0.33$ \\
$9-10$  & 192 & $2.99 \pm 0.16$ & 14 & $2.23 \pm 0.56$ \\
$10-11$ & 15  & $4.00 \pm 0.32$ & 11 & $2.36 \pm 0.48$ \\
\hline
\hline
\end{tabular}
\label{tab:r80_medians}
\tablecomments{Uncertainties on binned median values were estimated via non-parametric bootstrap resampling (2000 realizations per mass bin, sampling with replacement). We report the standard deviation of the bootstrap median distribution as the ($1\sigma$) error.}
\end{table}

\begin{figure}
    \centering
    \includegraphics[width = 0.7\textwidth]{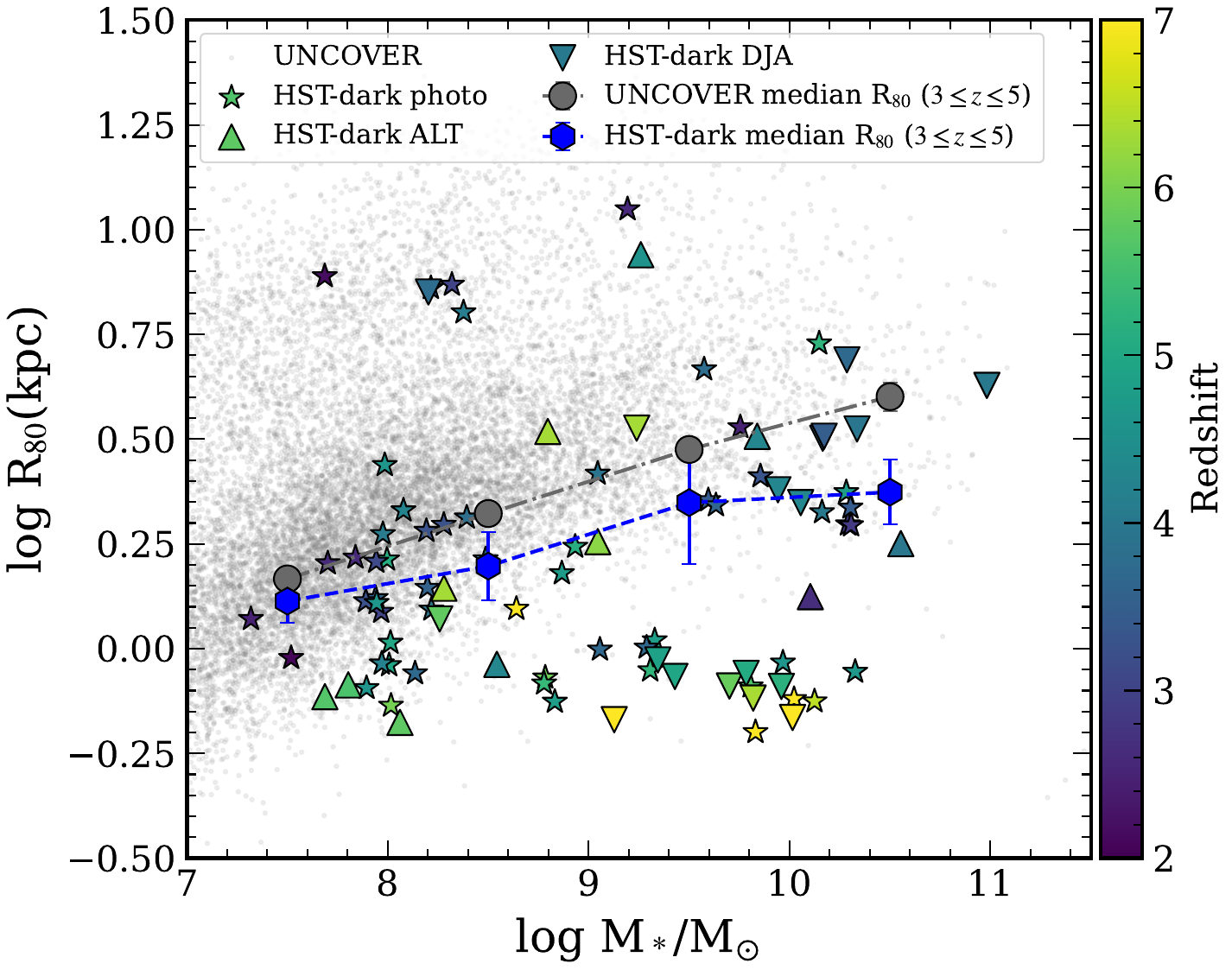}
    \caption{Size ($R_{80}$) plotted as a function of stellar mass. Grey points represent all the UNCOVER galaxies. Our sample of HST-dark galaxies is shown with stars and upper and lower triangles, colored by redshift. Median sizes in different stellar mass bins are also shown for the UNCOVER (grey circles) and HST-dark (blue hexagon) galaxies with $3\leq z \leq 5$. The error bars around the median values represent the standard deviations of the bootstrap median distribution with 2000 realizations. HST-dark galaxies are smaller in size than the typical UV-bright galaxies in all stellar mass bins.}
    \label{fig: Size_mass}
\end{figure}

To investigate the morphological nature of the HST-dark population, we perform a structural analysis using statmorph \citep{Rodriguez2019}, a Python-based code for computing non-parametric morphology parameters. We adopt the 80\% light radius ($r_{80}$, an output from statmorph) as our primary measure of galaxy size. We run statmorph on the  F444W filter cutouts of our galaxies and compare $r_{80}$ of our HST-dark sample against the broader UNCOVER population within the redshift range $3 < z < 5$. Our results reveal a clear and systematic offset in the Size–Mass relation over four orders of magnitude in stellar mass. HST-dark galaxies are consistently smaller in size than the UV-bright galaxies (see Figure \ref{fig: Size_mass}), and the difference in size increases with the stellar mass. 

At the lower-mass end ($\log M_{\star}/M_{\odot} = 7-8$), the median $r_{80}$ for the general UNCOVER population is $\sim 1.47$ kpc. In contrast, the HST-dark exhibit a smaller median size of $\sim 1.30$ kpc. This trend continues into the high-mass regime ($\log M_{\star}/M_{\odot} = 10-11$), where the difference becomes even more pronounced. HST-dark galaxies have median sizes of $\sim 2.36$ kpc, nearly 40\% smaller than the $\sim 4.00$ kpc median observed for the broader population in the same mass bin (see Table \ref{tab:r80_medians}).

The observed smaller sizes of these galaxies suggest that star formation in the HST-dark population is occurring in extremely dense environments. In a ``bottom-up" assembly scenario, these compact, dusty systems likely represent the rapid formation of stellar bulges. The high dust attenuation and small $r_{80}$ values indicate that these galaxies are efficiently converting gas into stars within centralized regions, potentially leading to the formation of the bulge components seen in modern-day massive galaxies.

Furthermore, the fact that this compactness is maintained down to the $10^{8} M_{\odot}$ mass regime suggests that the ``dark" nature of these galaxies is intrinsically tied to their structural density. In these compact halos, the gas and dust are concentrated enough to produce significant obscuration even at relatively low total dust masses. This structural evidence supports the idea that the HST-dark population represents a critical, high-density phase of galaxy evolution that is fundamentally different from the more extended, UV-bright star-forming galaxies typically identified at these redshifts.

\subsection{Cosmic Star Formation Rate Density (SFRD)} \label{subsec: SFRD}
 
Tracing the cosmic SFRD over time is fundamental to understanding the cosmic SFH \citep{Madau2014}, the galaxy mass assembly, and the metal enrichment history of the Universe. Historically, estimates of the SFRD at $z>3$ were heavily relied on UV-selected galaxies \citep{Bouwens2012a, Bouwens2012b}, predominantly Lyman-break galaxies (LBGs). However, as UV emission is highly susceptible to dust attenuation, these rest-frame UV samples are systematically biased against massive, dust-obscured, star-forming galaxies. Consequently, excluding these optically faint sources from star-formation censuses inevitably leads to a significant, redshift-dependent underestimation of the true total cosmic SFRD \citep{Cochrane2024}.

\begin{figure}
    \centering
    \includegraphics[width = 0.7\textwidth]{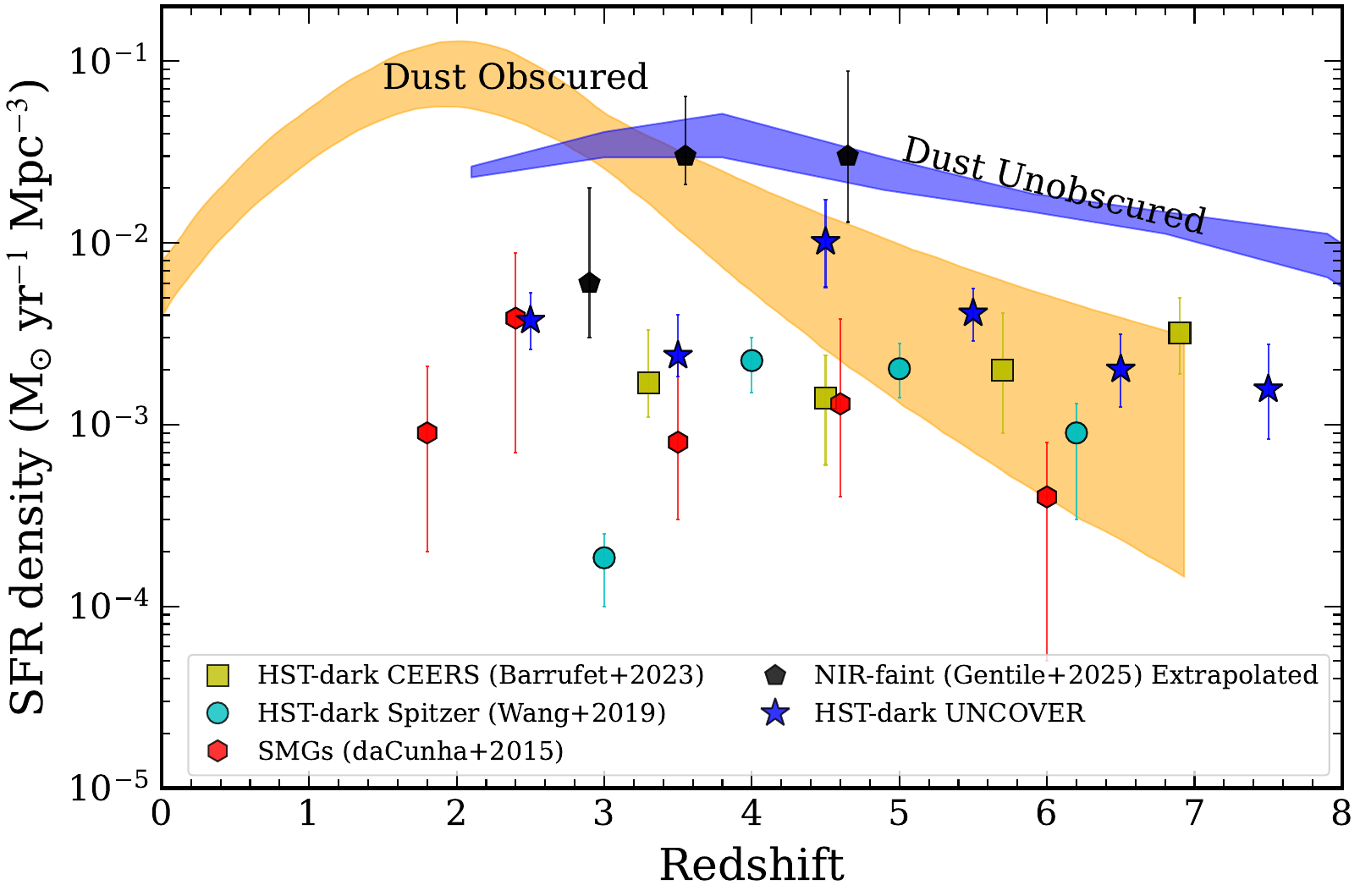}
    \caption{SFRD vs redshift is depicted for the HST-dark sample of ours (blue stars). SFRDs from various previous studies of SMGs and HST-dark galaxies are also shown. The orange area is the dust-obscured SFH derived from semi-empirical models by \citet{Zavala2021}. The dust-unobscured SFH is shown in blue, based on \citet{Bouwens2022}. We have extended the SFRD estimation for HST-dark galaxies to $z=8$. The SFRD estimated from our sample is slightly higher than that from previous optically dark galaxy samples, except in the redshift bin $4-5$, where it is significantly higher.}
    \label{fig: CSFRD}
\end{figure}

The study of HST-dark galaxies and SMGs has revealed that a significant portion of the cosmic SFH has been hidden. At $z=4-5$, \citet{Wang2016} estimated that these sources contribute about $15\%-25\%$ to the cosmic SFRD. \citet{Xiao2023} investigated the contribution of optically faint galaxies (OFGs) excluding LBGs to the cosmic SFRD and found that at $z>3$, massive OFGs contribute to cosmic SFRD at least two orders of magnitude higher than the similarly massive LBGs. They have also found that at $z=4-5$, the combined contribution of LBGs and OFGs is $43\%$ higher than the UV-selected samples to cosmic SFRD. Furthermore, JWST observations from CEERS have shown that the contribution of HST-dark galaxies to the cosmic SFRD remains relatively constant from $z\sim4$ all the way up to $z\sim7$ \citep{barrufet2023}. This is a critical finding because the contribution of classical submillimeter galaxies rapidly drops off at $z>5$ \citep{daCunha2015, Michalowski2017}. Extrapolations of radio-selected near-infrared-faint galaxies even suggest that their contribution to the total cosmic SFRD could equal that of UV-bright galaxies at $z\sim4$ \citep{Gentile2025}, confirming that tracing this population is absolutely pivotal for achieving a complete map of early galaxy evolution (see Figure \ref{fig: CSFRD}). 

\begin{table}
\centering
\caption{SFRD for the HST-dark Galaxy Sample.}
\begin{tabular}{c c c}
\hline
\hline
Redshift ($z$) & SFRD [$10^{-3} \, M_{\odot} \, \text{yr}^{-1} \, \text{Mpc}^{-3}$] & $N$ \\
\hline
2.5 & $3.75^{+1.55}_{-1.15}$ & 17 \\
3.5 & $2.40^{+1.64}_{-0.56}$ & 36 \\
4.5 & $10.14^{+7.10}_{-4.44}$ & 30 \\
5.5 & $4.10^{+1.52}_{-1.22}$ & 18 \\
6.5 & $2.02^{+1.12}_{-0.77}$ & 7 \\
7.5 & $1.57^{+1.20}_{-0.73}$ & 5 \\
\hline
\hline
\end{tabular}
\label{tab:sfrd_results}
\tablecomments{Errors represent the total asymmetric $1\sigma$ uncertainty, calculated by combining the Poissonian statistical errors of the galaxy counts and the individual SFR measurement uncertainties (from SED fitting) in quadrature.}
\end{table}

To quantify the contribution of our sample to the cosmic star formation budget, we calculate the cosmic SFRD ($\rho_{\text{SFR}}$) across six redshift bins spanning $z=2$ to $z=8$. The SFRD is determined by summing the individual galaxy SFRs within each redshift bin and normalizing by the comoving volume. The volume is calculated based on the survey area and a standard $\Lambda$CDM cosmology. Asymmetric $1\sigma$ uncertainties are derived by combining the Poissonian errors of the galaxy counts, following the frequentist approach for small sample sizes described by \citet{Gehrels1986}, with the individual SFR measurement uncertainties propagated in quadrature.

Our results, summarised in Table \ref{tab:sfrd_results}, reveal a significant and dynamic evolution of the obscured star formation density. We find that the cosmic SFRD peaks prominently at $z \approx 4.5$ with a value of $10.14^{+7.10}_{-4.44} \times 10^{-3} \, M_{\odot} \, \text{yr}^{-1} \, \text{Mpc}^{-3}$. This peak is notably higher than the cosmic SFRD observed in the adjacent bins ($z=3.5$ and $z=5.5$), suggesting that the epoch around $z \sim 4.5$ represents a period of intense, dust-shrouded galaxy assembly (see Figure \ref{fig: CSFRD}).

This finding provides an interesting comparison to the results of \citet{barrufet2023}. While their analysis of the CEERS field suggested a relatively constant contribution from HST-dark galaxies to the cosmic SFRD from $z \sim 4$ to $z \sim 7$, our sample—leveraging the increased depth of UNCOVER and the lensing magnification of the cluster—reveals a more punctuated evolution. Specifically, the sharp increase at $z \sim 4.5$ suggests that the ``optically dark'' phase may be more prevalent at specific cosmic epochs than previously recognized, potentially driven by the rapid evolution and dust-enrichment of the lower-mass galaxy population ($\log M_{\star}/M_{\odot} \approx 9$) that dominates our sample.

Furthermore, our data confirm the presence of obscured star formation into the Epoch of Reionization. At $z=6.5$ and $z=7.5$, we find SFRD values of $2.02 \times 10^{-3}$ and $1.57 \times 10^{-3} \, M_{\odot} \, \text{yr}^{-1} \, \text{Mpc}^{-3}$, respectively. While the contribution of classical, massive SMGs is known to drop precipitously beyond $z > 5$ \citep{daCunha2015}, our detection of these intrinsically fainter HST-dark sources indicates that dust-obscured star formation remains a non-negligible component of the total stellar mass assembly budget even at these early times (see Figure \ref{fig: CSFRD}). This reinforces the necessity of additional ultra-deep JWST observations to capture the full star-forming census, as these ``dusty dwarfs'' provide a vital link between the early UV-bright population and the massive, obscured systems observed at cosmic noon. However, we caution that, as our analysis is confined to a single JWST field, the observed cosmic SFRD evolution, particularly the sharp peak at $z \approx 4.5$, may be affected by cosmic variance or localized overdensities. Expanding this study across multiple independent deep JWST surveys will be crucial to averaging out these effects and establishing a truly global census of this obscured population.

\section{Summary and Conclusion} \label{sec: Summary}

Here, we present a comprehensive analysis of 113 high-redshift ($z > 2$) and 94 low-redshift ($z < 2$) HST-dark galaxies identified within the ultra-deep JWST UNCOVER and MegaScience surveys of the Abell 2744 field. By leveraging the unprecedented sensitivity of NIRCam broad-band and medium-band imaging, along with the magnification provided by the foreground cluster, we have probed a previously hidden population of galaxies. Our findings challenge the traditional view that the ``optically dark" universe is exclusively populated by massive, star-forming galaxies. The primary conclusions of this work are summarised below.

\begin{itemize}
    \item Our primary sample of HST-dark galaxies at $z > 2$ reaches the ``dusty dwarf" regime with a median stellar mass of $\log M_{\star}/M_{\odot} \approx 8.8$, unlike previous \textit{Spitzer}-selected samples that were limited to the high-mass regime ($\log M_{\star}/M_{\odot} \gtrsim 10$). This demonstrates that significant dust obscuration is a prevalent feature across a wide range of stellar masses at high redshift.

    \item We find a median dust attenuation of $A_V = 0.99$, which is significantly lower than the values reported for classical SMGs or more massive HST-dark samples ($A_V \gtrsim 2.0$). This reveals that these galaxies are ``dark'' to Hubble, not necessarily due to extreme dust columns, but because their intrinsic faintness (low mass and low SFR) combined with moderate dust, which is sufficient to suppress their rest-frame UV and optical flux below the HST detection limit.

    \item Our sample sits squarely on the Star-Forming Main Sequence. This indicates that HST-dark galaxies are not short-lived, exotic outliers but rather represent a typical yet obscured phase of galaxy assembly.

    \item Size-mass analysis reveals that HST-dark galaxies are systematically smaller in size than the UV-bright galaxies across all mass bins. Their lower $r_{80}$ values suggest that star formation in these sources occurs in high-density environments, making them potential progenitors of the central bulges seen in massive galaxies today.

    \item Our calculated SFRD shows a prominent peak at $z \approx 4.5$ ($10.14^{+7.10}_{-4.44} \times 10^{-3} \, M_{\odot} \, \text{yr}^{-1} \, \text{Mpc}^{-3}$). The persistence of these obscured galaxies into the Epoch of Reionization ($z > 6$) confirms that a substantial portion of the early stellar mass assembly has been systematically underestimated in UV-only surveys.

    \item We also identified a distinct sub-population of optically dark sources at $z < 2$ (median $z \approx 0.76$) that exhibit properties remarkably similar to the HELM galaxies recently identified and characterized with JWST. In our sample, these galaxies are extremely low-mass (median $\log M_{\star}/M_{\odot} \approx 7.10$) and disproportionately dusty, with a median attenuation of $A_V \approx 1.97$ mag, which is significantly higher than that of our high-redshift sample. These galaxies are characterised by extremely low SFRs (median $\text{SFR} \approx 0.009 \, M_{\odot} \text{yr}^{-1}$) and broad photometric redshift posterior distributions.
\end{itemize}

The diversity of our HST-dark sample, ranging from compact, massive star-forming galaxies at cosmic noon to extremely dusty dwarfs at $z < 1$, underscores the complexity of dust enrichment and galaxy assembly across cosmic time. Our findings, particularly the prominent peak in the SFRD at $z \approx 4.5$, suggest that a substantial portion of the dust-obscured star-formation in the early universe has been missed by previous surveys. Collectively, our results suggest that dust production occurs rapidly and efficiently even in low-mass galaxies, and that the transition to a dust-obscured interstellar medium is a fundamental evolutionary stage that begins earlier and at lower mass scales than previously seen. To mitigate the impact of cosmic variance and confirm the global significance of these findings, it will be imperative to extend such ultra-deep multi-band observations to additional JWST surveys. Future spectroscopic follow-up with JWST/NIRSpec will also be essential to confirm the photometric redshifts and characterize the metallicities and gas-phase kinematics of these galaxies.  

\begin{acknowledgements}
We thank the anonymous referee whose insightful comments improved both the content and presentation of this paper.
The authors acknowledge the support of the Department of Atomic Energy, Government of India, under the project title ``Next Generation Instrumentation for Radio Astronomy'', UID number RTI4017.
We are grateful to the UNCOVER and MegaScience teams for designing their observing programs, developing data-reduction pipelines, and providing reduced data products to the public. 
This work is based on observations made with the NASA/ESA/CSA James Webb Space Telescope and the NASA/ESA Hubble Space Telescope. The data used in this paper were obtained from the JWST-UNCOVER website\footnote{\url{https://jwst-uncover.github.io/\#releases}}. These observations are associated with JWST Cycle 1 GO program 2561 and Cycle 2 GO program 4111, and gratefully made use of additional public JWST programs in the Abell 2744 field, including JWST-ERS-1324, JWST-DD-2756, JWST-GO-2641, JWST-GO-2883, JWST-GO-3516, and JWST-GO-3538. The relevant HST observations are associated with programs HST-GO-11689, HST-GO-13386, HST-GO-13389, HST-GO/DD-13495, HST-GO-15117, and HST-GO/DD-17231.
Some of the data products presented herein were retrieved from the Dawn JWST Archive (DJA). DJA is an initiative of the Cosmic Dawn Center (DAWN), which is funded by the Danish National Research Foundation under grant DNRF140.
\end{acknowledgements}

\facilities{JWST, HST}

\software{astropy \citep{Astropy2013, Astropy2018, astropy2022},
          numpy \citep{numpy}, 
          matplotlib \citep{matplotlib},
          pandas \citep{pandas2010, pandas2020}, 
          ds9 \citep{ds91, ds92, ds93, ds94, ds95},
          TOPCAT \citep{TOPCAT2005}
          }

\appendix
\section{Mean posterior distributions of redshift and stellar mass of the $z>2$ HST-dark galaxies} \label{sec: posterior}
\begin{figure}
    \centering
    \includegraphics[width = 0.9\textwidth]{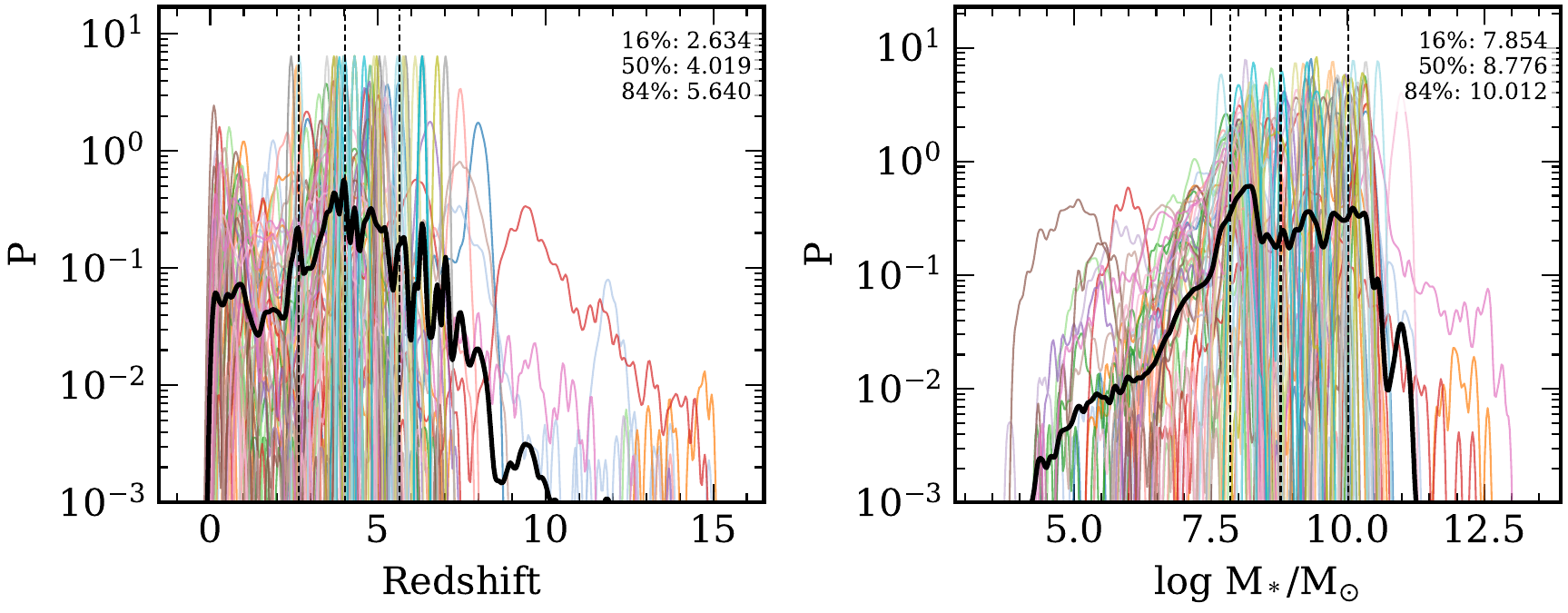}
    \caption{Redshift (left panel) and stellar mass (right panel) posterior distributions of the entire $z>2$ HST-dark galaxy sample (colored lines). The average posteriors are shown as thick, solid black lines. The 16th, 50th, and 84th percentiles of the redshift and stellar mass, calculated from the mean posteriors (solid black lines), are shown in the upper-right corners.}
    \label{fig: posterior}
\end{figure}

\texttt{BAGPIPES} provides full posterior distributions for all fitted parameters for each galaxy. In Figure \ref{fig: posterior}, we show the redshift and stellar mass posteriors for all $z>2$ HST-dark galaxies, along with their mean posteriors (thick black lines) and corresponding percentiles. The 50th-percentile values for both redshift and stellar mass closely match the medians of the $z>2$ HST-dark galaxies. This implies that the median values of the output 1D posteriors derived from BAGPIPES are similar to the median of the averaged posteriors for stellar mass and redshift.

\section{HST-dark sample ($z>2$)} \label{sec: sample_high}
See Table \ref{tab:highz_sample} for the properties of the HST-dark galaxies ($z>2$). The spectroscopic samples from ALT and DJA are given in Tables \ref{tab:alt_sample} and \ref{tab:dja_sample}. For more information, see Sections \ref{sec: Data} and \ref{sec: SPS} for our sample selection and SPS fitting with \texttt{BAGPIPES}.

\begin{table}
\centering
\caption{Properties of HST-dark galaxies ($z>2$)} \label{sec: sample_high}
\begin{tabular}{lccccccccccc}
\hline
ID\_DR3 & ra & dec & Redshift & $\log \frac{M_\star}{M_\odot}$ & $\log \mathrm{SFR}$ & U--V & V--J & Av & r80 & $\mu$ \\
& (deg) & (deg) & & & ($\rm M_{\odot}yr^{-1}$) & (mag) & (mag) & (mag) & (arcsec) & \\
\hline
508 & 3.6153722 & -30.4658274 & $5.62^{+0.13}_{-0.07}$ & $9.81^{+0.06}_{-0.08}$ & $1.30^{+0.10}_{-0.10}$ & $0.95^{+0.03}_{-0.03}$ & $0.48^{+0.06}_{-0.05}$ & $0.96^{+0.09}_{-0.09}$ & 0.13 & 1.27 \\
1171 & 3.6169204 & -30.4630031 & $5.27^{+0.09}_{-0.08}$ & $9.31^{+0.21}_{-0.12}$ & $1.50^{+0.34}_{-0.14}$ & $0.86^{+0.11}_{-0.09}$ & $1.22^{+0.08}_{-0.09}$ & $1.40^{+0.31}_{-0.15}$ & 0.14 & 1.28 \\
2506 & 3.6270847 & -30.4578844 & $7.94^{+0.23}_{-0.31}$ & $9.83^{+0.15}_{-0.13}$ & $1.76^{+0.29}_{-0.25}$ & $1.38^{+0.09}_{-0.07}$ & $1.47^{+0.31}_{-0.25}$ & $2.09^{+0.33}_{-0.27}$ & 0.13 & 1.28 \\
4634 & 3.6230042 & -30.4500143 & $2.95^{+0.23}_{-0.20}$ & $10.29^{+0.12}_{-0.17}$ & $2.30^{+0.24}_{-0.50}$ & $1.74^{+0.13}_{-0.11}$ & $2.12^{+0.14}_{-0.10}$ & $3.00^{+0.20}_{-0.50}$ & 0.25 & 1.27 \\
6954 & 3.5946059 & -30.4420220 & $4.78^{+0.03}_{-0.05}$ & $9.33^{+0.03}_{-0.03}$ & $1.64^{+0.03}_{-0.04}$ & $0.70^{+0.03}_{-0.03}$ & $0.86^{+0.05}_{-0.07}$ & $1.24^{+0.04}_{-0.05}$ & 0.16 & 1.51 \\
7867 & 3.5691162 & -30.4386690 & $3.51^{+0.58}_{-2.35}$ & $7.94^{+0.17}_{-0.57}$ & $-0.61^{+0.24}_{-0.89}$ & $0.51^{+0.45}_{-0.19}$ & $0.12^{+0.59}_{-0.21}$ & $0.33^{+0.47}_{-0.18}$ & 0.18 & 1.40 \\
8128 & 3.5671172 & -30.4380087 & $3.81^{+0.24}_{-0.43}$ & $8.39^{+0.10}_{-0.14}$ & $-0.20^{+0.16}_{-0.16}$ & $0.39^{+0.12}_{-0.11}$ & $-0.12^{+0.18}_{-0.11}$ & $0.15^{+0.15}_{-0.11}$ & 0.28 & 1.41 \\
9248 & 3.6207043 & -30.4351901 & $2.49^{+2.02}_{-0.64}$ & $9.75^{+0.57}_{-0.40}$ & $1.01^{+1.05}_{-0.51}$ & $2.79^{+0.58}_{-0.79}$ & $3.25^{+0.73}_{-1.00}$ & $4.49^{+0.96}_{-1.33}$ & 0.41 & 1.43 \\
10708 & 3.6439445 & -30.4314827 & $3.45^{+0.21}_{-1.62}$ & $9.60^{+0.15}_{-0.29}$ & $1.06^{+0.37}_{-0.61}$ & $1.33^{+0.68}_{-0.14}$ & $1.21^{+0.66}_{-0.16}$ & $1.72^{+0.92}_{-0.34}$ & 0.30 & 1.30 \\
10729 & 3.5864273 & -30.4313678 & $3.56^{+0.11}_{-0.18}$ & $8.20^{+0.06}_{-0.10}$ & $-0.41^{+0.12}_{-0.08}$ & $0.32^{+0.06}_{-0.07}$ & $-0.26^{+0.05}_{-0.04}$ & $0.07^{+0.08}_{-0.05}$ & 0.19 & 1.31 \\
\hline
\end{tabular}
\label{tab:highz_sample}
\tablecomments{The full table is available in a machine-readable form in the online article.}
\end{table}

\begin{table}
\centering
\caption{Properties of HST-dark galaxies ($z>2$) with spectroscopic redshift from ALT survey}
\begin{tabular}{lccccccccccc}
\hline
ID\_DR3 & ra & dec & Redshift & $\log \frac{M_\star}{M_\odot}$ & $\log \mathrm{SFR}$ & U--V & V--J & Av & r80 & $\mu$ \\
& (deg) & (deg) & & & ($\rm M_{\odot}yr^{-1}$) & (mag) & (mag) & (mag) & (arcsec) & \\
\hline
11682 & 3.5675211 & -30.4287359 & 5.76 & $8.06^{+0.14}_{-0.21}$ & $-0.20^{+0.17}_{-0.14}$ & $0.49^{+0.11}_{-0.10}$ & $-0.01^{+0.13}_{-0.13}$ & $0.33^{+0.13}_{-0.13}$ & 0.11 & 1.52 \\
12602 & 3.5894544 & -30.4261916 & 6.12 & $9.05^{+0.04}_{-0.03}$ & $1.38^{+0.03}_{-0.03}$ & $0.81^{+0.04}_{-0.04}$ & $0.60^{+0.05}_{-0.05}$ & $0.99^{+0.05}_{-0.04}$ & 0.31 & 1.99 \\
13415 & 3.5754719 & -30.4244692 & 4.02 & $9.43^{+0.08}_{-0.08}$ & $1.19^{+0.22}_{-0.14}$ & $1.23^{+0.04}_{-0.04}$ & $1.42^{+0.09}_{-0.06}$ & $1.67^{+0.25}_{-0.10}$ & 0.78 & 1.78 \\
15465 & 3.5759869 & -30.4190294 & 3.83 & $8.01^{+0.08}_{-0.07}$ & $0.28^{+0.07}_{-0.07}$ & $0.85^{+0.07}_{-0.08}$ & $1.34^{+0.17}_{-0.14}$ & $1.31^{+0.12}_{-0.12}$ & 0.11 & 1.90 \\
15610 & 3.5987977 & -30.4187332 & 4.32 & $8.54^{+0.30}_{-0.04}$ & $0.88^{+0.31}_{-0.04}$ & $0.52^{+0.24}_{-0.05}$ & $0.99^{+0.05}_{-0.06}$ & $0.99^{+0.32}_{-0.06}$ & 0.13 & 2.69 \\
21839 & 3.6124251 & -30.4056651 & 6.33 & $8.80^{+0.05}_{-0.05}$ & $1.06^{+0.05}_{-0.06}$ & $0.80^{+0.06}_{-0.06}$ & $0.86^{+0.10}_{-0.07}$ & $1.14^{+0.06}_{-0.06}$ & 0.58 & 2.01 \\
21888 & 3.6112710 & -30.4054001 & 6.32 & $8.28^{+0.04}_{-0.03}$ & $0.62^{+0.04}_{-0.04}$ & $0.55^{+0.03}_{-0.03}$ & $0.06^{+0.06}_{-0.05}$ & $0.52^{+0.04}_{-0.04}$ & 0.25 & 2.10 \\
27716 & 3.6175093 & -30.3953555 & 4.58 & $9.26^{+0.05}_{-0.03}$ & $1.53^{+0.04}_{-0.07}$ & $0.71^{+0.06}_{-0.06}$ & $0.74^{+0.05}_{-0.04}$ & $1.05^{+0.04}_{-0.04}$ & 1.30 & 1.65 \\
35257 & 3.5353067 & -30.3810088 & 5.66 & $7.69^{+0.06}_{-0.05}$ & $0.03^{+0.05}_{-0.05}$ & $0.61^{+0.05}_{-0.05}$ & $0.18^{+0.09}_{-0.07}$ & $0.63^{+0.07}_{-0.06}$ & 0.13 & 2.57 \\
35891 & 3.5920965 & -30.3804717 & 2.65 & $10.10^{+0.03}_{-0.02}$ & $2.42^{+0.02}_{-0.03}$ & $1.72^{+0.04}_{-0.04}$ & $2.87^{+0.03}_{-0.03}$ & $3.12^{+0.05}_{-0.05}$ & 0.16 & 2.35 \\
\hline
\end{tabular}
\label{tab:alt_sample}
\tablecomments{The full table is available in a machine-readable form in the online article.}
\end{table}

\begin{table}
\centering
\caption{Properties of HST-dark galaxies ($z>2$) with spectroscopic redshift from DJA}
\begin{tabular}{lccccccccccc}
\hline
ID\_DR3 & ra & dec & Redshift & $\log \frac{M_\star}{M_\odot}$ & $\log \mathrm{SFR}$ & U--V & V--J & Av & r80 & $\mu$ \\
& (deg) & (deg) & & & ($\rm M_{\odot}yr^{-1}$) & (mag) & (mag) & (mag) & (arcsec) & \\
\hline
13416 & 3.5755647 & -30.4243802 & 4.02 & $10.98^{+0.10}_{-0.12}$ & $2.96^{+0.35}_{-0.56}$ & $1.92^{+0.06}_{-0.04}$ & $2.35^{+0.22}_{-0.17}$ & $3.32^{+0.31}_{-0.51}$ & 0.60 & 1.78 \\
13742 & 3.6192037 & -30.4232716 & 5.84 & $9.70^{+0.11}_{-0.11}$ & $2.03^{+0.11}_{-0.11}$ & $1.08^{+0.08}_{-0.07}$ & $1.36^{+0.16}_{-0.13}$ & $1.86^{+0.17}_{-0.15}$ & 0.14 & 1.61 \\
20698 & 3.5567049 & -30.4081918 & 2.42 & $9.37^{+0.07}_{-0.05}$ & $1.60^{+0.04}_{-0.05}$ & $1.29^{+0.05}_{-0.05}$ & $1.80^{+0.03}_{-0.03}$ & $1.99^{+0.08}_{-0.06}$ & 0.14 & 1.54 \\
21547 & 3.5508378 & -30.4065978 & 5.06 & $9.78^{+0.05}_{-0.08}$ & $2.12^{+0.06}_{-0.08}$ & $1.04^{+0.03}_{-0.04}$ & $1.25^{+0.08}_{-0.06}$ & $1.75^{+0.07}_{-0.09}$ & 0.14 & 1.61 \\
21838 & 3.6123396 & -30.4056967 & 6.33 & $9.24^{+0.07}_{-0.06}$ & $1.35^{+0.10}_{-0.09}$ & $0.85^{+0.06}_{-0.06}$ & $0.90^{+0.10}_{-0.09}$ & $1.18^{+0.10}_{-0.08}$ & 0.59 & 2.01 \\
24175 & 3.5798316 & -30.4015692 & 7.04 & $10.01^{+0.09}_{-0.08}$ & $1.78^{+0.22}_{-0.16}$ & $1.87^{+0.05}_{-0.05}$ & $2.00^{+0.17}_{-0.13}$ & $2.96^{+0.20}_{-0.17}$ & 0.13 & 6.87 \\
24968 & 3.6206068 & -30.3999507 & 6.35 & $9.82^{+0.06}_{-0.06}$ & $2.17^{+0.06}_{-0.06}$ & $1.36^{+0.06}_{-0.06}$ & $2.10^{+0.09}_{-0.09}$ & $2.24^{+0.11}_{-0.12}$ & 0.13 & 1.62 \\
27114 & 3.5457903 & -30.3957238 & 5.20 & $9.96^{+0.14}_{-0.21}$ & $2.16^{+0.22}_{-0.30}$ & $1.74^{+0.07}_{-0.07}$ & $2.33^{+0.13}_{-0.12}$ & $3.15^{+0.23}_{-0.38}$ & 0.13 & 1.84 \\
27966 & 3.5972031 & -30.3943301 & 7.03 & $9.13^{+0.11}_{-0.09}$ & $1.38^{+0.14}_{-0.14}$ & $1.34^{+0.06}_{-0.06}$ & $1.53^{+0.16}_{-0.14}$ & $2.03^{+0.19}_{-0.15}$ & 0.13 & 3.60 \\
28380 & 3.6326605 & -30.3936353 & 3.96 & $10.16^{+0.06}_{-0.09}$ & $1.68^{+0.26}_{-0.14}$ & $1.16^{+0.04}_{-0.04}$ & $1.15^{+0.09}_{-0.08}$ & $1.39^{+0.25}_{-0.18}$ & 0.45 & 1.40 \\
\hline
\hline
\end{tabular}
\label{tab:dja_sample}
\tablecomments{The full table is available in a machine-readable form in the online article.}
\end{table}

\section{HST-dark sample ($z<2$)} \label{sec: sample_low}
 
Table~\ref{tab:lowz_sample} represents the properties of the HST-dark galaxies with $z<2$. Also see section \ref{sec: Lowz}, where we have discussed these low-redshift HST-dark galaxies.

\begin{table}
\centering
\caption{Properties of HST-dark galaxies ($z<2$)}
\begin{tabular}{lcccccccccc}
\hline
ID\_DR3 & ra & dec & Redshift & $\log \frac{M_\star}{M_\odot}$ & $\log \mathrm{SFR}$ & U--V & V--J & Av & r80 & $\mu$ \\
& (deg) & (deg) & & & ($\rm M_{\odot}yr^{-1}$) & (mag) & (mag) & (mag) & (arcsec) & \\
\hline
987 & 3.6556036 & -30.4637311 & $0.84^{+1.34}_{-0.33}$ & $6.99^{+0.55}_{-0.47}$ & $-2.17^{+0.98}_{-0.84}$ & $1.87^{+1.17}_{-0.96}$ & $1.68^{+1.56}_{-1.10}$ & $2.17^{+2.17}_{-1.42}$ & 0.05 & 1.20 \\
1361 & 3.6528633 & -30.4623134 & $0.12^{+0.04}_{-0.05}$ & $5.50^{+0.56}_{-0.44}$ & $-3.37^{+0.36}_{-0.40}$ & $2.83^{+0.61}_{-0.97}$ & $2.98^{+0.84}_{-1.22}$ & $4.44^{+1.13}_{-1.85}$ & 0.23 & 1.22 \\
1456 & 3.5997845 & -30.4618045 & $0.83^{+2.50}_{-0.24}$ & $7.26^{+0.70}_{-0.35}$ & $-1.88^{+1.29}_{-0.53}$ & $1.52^{+0.63}_{-0.96}$ & $1.22^{+0.76}_{-1.10}$ & $1.47^{+1.11}_{-1.08}$ & 0.26 & 1.29 \\
2056 & 3.6061578 & -30.4597327 & $0.78^{+2.94}_{-0.47}$ & $7.07^{+0.75}_{-0.74}$ & $-2.09^{+1.50}_{-1.64}$ & $1.28^{+1.26}_{-0.90}$ & $0.98^{+1.46}_{-0.99}$ & $1.06^{+1.94}_{-0.81}$ & 0.44 & 1.30 \\
2180 & 3.5982979 & -30.4592092 & $0.59^{+1.07}_{-0.31}$ & $7.32^{+0.69}_{-0.70}$ & $-2.02^{+1.22}_{-0.83}$ & $2.89^{+0.69}_{-1.38}$ & $3.10^{+0.89}_{-1.82}$ & $4.09^{+1.29}_{-2.34}$ & 0.26 & 1.33 \\
2188 & 3.6316643 & -30.4591923 & $0.69^{+1.53}_{-0.28}$ & $7.31^{+0.67}_{-0.44}$ & $-1.86^{+1.18}_{-0.80}$ & $1.36^{+0.91}_{-0.75}$ & $1.05^{+0.99}_{-0.82}$ & $1.14^{+1.31}_{-0.83}$ & 0.20 & 1.26 \\
2476 & 3.6545657 & -30.4579697 & $0.10^{+0.07}_{-0.05}$ & $5.34^{+0.67}_{-0.55}$ & $-3.75^{+0.50}_{-0.57}$ & $2.10^{+0.93}_{-0.98}$ & $1.92^{+1.35}_{-1.15}$ & $2.59^{+1.82}_{-1.50}$ & 0.11 & 1.22 \\
2772 & 3.5949458 & -30.4573007 & $0.28^{+0.01}_{-0.01}$ & $6.77^{+0.11}_{-0.11}$ & $-2.19^{+0.09}_{-0.09}$ & $3.58^{+0.14}_{-0.23}$ & $4.31^{+0.20}_{-0.21}$ & $5.71^{+0.20}_{-0.34}$ & 1.11 & 1.32 \\
2776 & 3.5946321 & -30.4570628 & $1.82^{+0.42}_{-0.26}$ & $9.92^{+0.14}_{-0.12}$ & $1.02^{+0.24}_{-0.20}$ & $2.22^{+0.21}_{-0.29}$ & $2.17^{+0.25}_{-0.28}$ & $2.98^{+0.72}_{-0.52}$ & 1.30 & 1.30 \\
3227 & 3.6089411 & -30.4550166 & $0.85^{+1.04}_{-0.20}$ & $7.30^{+0.54}_{-0.32}$ & $-1.79^{+0.86}_{-0.44}$ & $2.63^{+0.75}_{-1.18}$ & $2.78^{+0.97}_{-1.58}$ & $3.87^{+1.43}_{-2.31}$ & 0.14 & 1.33 \\
\hline
\end{tabular}
\label{tab:lowz_sample}
\tablecomments{The full table is available in a machine-readable form in the online article.}
\end{table}

\section{Spectrum of one quiescent galaxy} \label{sec: spectrum}
\begin{figure}
    \centering
    \includegraphics[width = 0.8\textwidth]{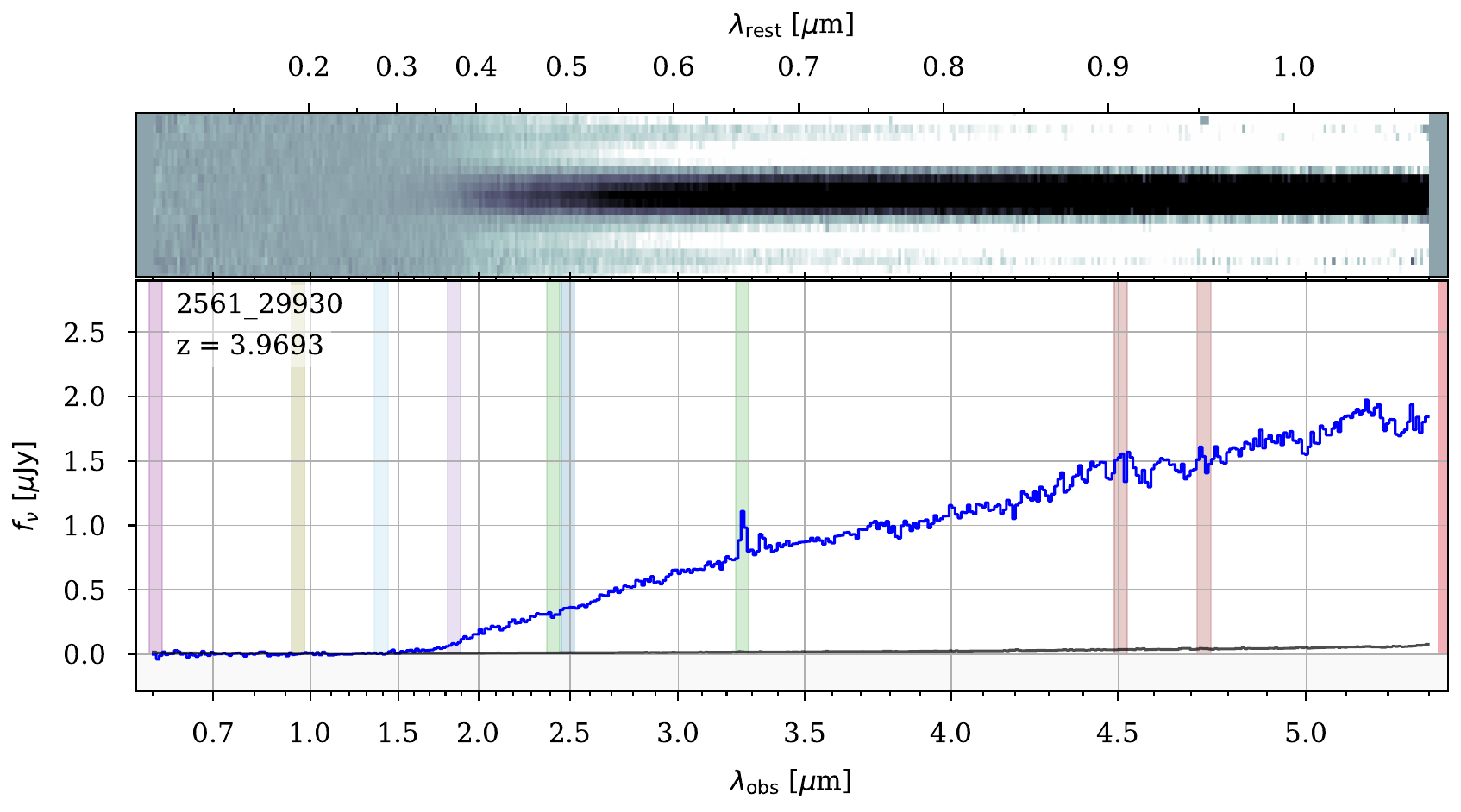}
    \caption{2D and 1D spectrum of one of the quenched galaxies in our sample at $z=3.96$. The positions of the multiple well-known lines are marked with color strips. We can see the red continuum nature of the galaxy. Also, the spectrum is devoid of emission lines apart from a very weak $\rm H\alpha$ line, as is expected from a quenched galaxy. }
    \label{fig: Spectrum}
\end{figure}

In Section \ref{subsec: Dust}, we have discussed the UVJ diagram diagnostic of the HST-dark galaxies, where we find 3 quiescent galaxies. The spectrum of one of the quiescent galaxies is shown in Figure \ref{fig: Spectrum}.

\bibliography{manuscript}{}
\bibliographystyle{aasjournalv7}

\end{document}